\documentclass[twocolumn,superscriptaddress,prl]{revtex4}

\usepackage[latin9]{inputenc}
\usepackage{color}
\usepackage{verbatim}
\usepackage{amsmath}
\usepackage{amssymb}
\usepackage{graphicx}
\usepackage{relsize}
\usepackage{esint}
\usepackage{multirow}
\usepackage{hyperref}

\makeatletter
\@ifundefined{textcolor}{}
{%
 \definecolor{BLACK}{gray}{0}
 \definecolor{WHITE}{gray}{1}
 \definecolor{RED}{rgb}{1,0,0}
 \definecolor{GREEN}{rgb}{0,1,0}
 \definecolor{BLUE}{rgb}{0,0,1}
 \definecolor{CYAN}{cmyk}{1,0,0,0}
 \definecolor{MAGENTA}{cmyk}{0,1,0,0}
 \definecolor{YELLOW}{cmyk}{0,0,1,0}
}

\usepackage{mathrsfs}

\newcommand{\beginsupplement}{%
        \setcounter{table}{0}
        \renewcommand{\thetable}{S\arabic{table}}%
        \setcounter{figure}{0}
        \renewcommand{\thefigure}{S\arabic{figure}}%
        \setcounter{equation}{0}
        \renewcommand{\theequation}{S\arabic{equation}}%
        \setcounter{section}{0}
        \renewcommand{\thesection}{S-\Roman{section}}%
     }
\draft

\begin{document}
\title{Quantum phase transition of fracton topological orders}
\author{Ting Fung Jeffrey Poon}
\affiliation{International Center for Quantum Materials, School of Physics, Peking University, Beijing 100871, China}
\affiliation{Collaborative Innovation Center of Quantum Matter, Beijing 100871, China}
\author{Xiong-Jun Liu
\footnote{Corresponding author: xiongjunliu@pku.edu.cn}}
\affiliation{International Center for Quantum Materials, School of Physics, Peking University, Beijing 100871, China}
\affiliation{Collaborative Innovation Center of Quantum Matter, Beijing 100871, China}
\affiliation{CAS Center for Excellence in Topological Quantum Computation, University of Chinese Academy of Sciences, Beijing 100190, China}
\affiliation{Institute for Quantum Science and Engineering and Department of Physics, Southern University of Science and Technology, Shenzhen 518055, China}

\begin{abstract}
Fracton topological order (FTO) is a new classification of correlated phases in three spatial dimensions with topological ground state degeneracy (GSD) scaling up with system size, and fractional excitations which are immobile or have restricted mobility. With the topological origin of GSD, FTO is immune to local perturbations, whereas a strong enough global external perturbation is expected to break the order. The critical point of the topological transition is however very challenging to identify. 
In this work, we propose to characterize quantum phase transition of the type-I FTOs induced by external terms and develop a theory to study analytically the critical point of the transition. 
In particular, for the external perturbation term creating lineon-type excitations, we predict a generic formula for the critical point of the quantum phase transition, characterized by the breaking-down of GSD. This theory applies to a board class of FTOs, including X-cube model, and for more generic FTO models under perturbations creating two-dimensional (2D) or 3D excitations, we predict the upper and lower limits of the critical point. Our work makes a step in characterizing analytically the quantum phase transition of generic fracton orders. 
\end{abstract}

\maketitle
\indent
\section{Introduction}
The discovery of topological quantum phases revolutionized the characterization and classification of fundamental states of quantum matter beyond the Landau symmetry-breaking paradigm. The extensive studies have brought about two broad categories of topological matter, the
symmetry-protected topological phases~\cite{Schnyder2008,Kitaev2009,Ryu2010,XieChen2012} and topologically ordered phases~\cite{xg89,xg90a,xg90b}, which exhibit short- and long-range quantum entanglement, respectively. Unlike the topological phases which necessitate symmetry-protection, the topological orders are characterized by finite ground state degeneracy (GSD) which is protected by the bulk gap and robust against any local perturbations, and host freely-moving fractional excitations~\cite{kitaevtc2,FQHE,FQHE2,NRead}. The topological protection of the GSD can have potential application to fault tolerant quantum computation~\cite{kitaevtc, hbmamd, dsarma08}, particularly for the systems which host non-Abelian excitations~\cite{non-Abelian1,non-Abelian2,non-Abelian3,non-Abelian4,non-Abelian5}.

Recently, a new basic class of topological phases, called fracton topological order (FTO)~\cite{Chamon, Sergey, Haah, Yoshida, Checkerboard, SpinChain, CoupledLayer, Regnault, Cagenet} were proposed and deeply broaden the understanding of the correlated topological states. In contrast to the conventional topological-ordered phases, the GSD of FTO scales up with the system size, and the fractional excitations of a FTO are immobile or have restricted mobility~\cite{Williamson, Parame, Regnault2, YMLu}. In particular, two types of FTOs are proposed and distinguished by their excitations~\cite{Checkerboard}. For type-I fracton orders, there are three categories of excitations, i.e. the 1D lineon excitations, 2D and 3D fractons, which are generated by 1D line, 2D membrane, and 3D body operators, respectively. The fractons are located at the ends of the line operator or corners of the membrane and body operators, and can move only along the line or in the way that the number of corners of the membrane/body operator remains the same. On the other hand, for the type-II fracton phases the excitations are located at the corners of a fractal operator and are completely immobile~\cite{Haah}. The restricted mobility of excitations is important for the stability of FTOs at finite temperature \cite{Haah2011, Haah2013} compared with conventional topological orders, and may have advantageous in applying to topological quantum computation.

The exploration of the FTOs is still in the relatively early stage. Efforts have been made in relating the fracton orders to other novel physics \cite{Gravity, GlassyD, QHE}, improving the basic concepts and characterizations of FTOs~\cite{QFTXCube, PRX3DMan,FoliatedFT}, and building new models for realization~\cite{Cagenet, QFTXCube}. So far the studies are performed on the fine-tuned toy models, which are exactly solvable but hard to achieve in the real physical systems. On the other hand, with the protection by the topological bulk gap, the FTOs are expected to be stable against small external perturbations and may undergo quantum phase transition when the system is tuned far away from the fine-tuning point. To uncover the critical point of such transition is crucial to determine the phase diagram, while very challenging in general, and may provide insights into the possible realization of FTOs in the real physical systems. This therefore brings a challenging but highly nontrivial issue to the frontier of this research direction, namely, characterization of the quantum phase transition of FTOs, 
as we study in this work.

In this article, we aim to characterize quantum phase transition in type-I FTOs induced by external perturbation terms, with a generic formula to obtain the critical point for the phase transition. The GSD is unchanged when the strength of external term is less than the critical value, and breaks down when the phase transition occurs as the external term exceeds the critical value. The universal results of the critical points are obtained. The article is organized as follow. In section II we briefly review the basics of FTOs. Then in section III we define the quantum phase transition in FTOs and present the generic formalism. The analytic theory of the critical point of phase transition is presented in section IV, with a universal result being obtained by mapping the present framework to a random walk theory and applicable to a class of FTOs, including the X-cube model. In section V, we generalize the present study based on random walk theory to more generic FTOs, and show the upper limits of the critical points of the phase transition. A reasonable conjecture with numerical verification of the lower limit of the critical points is also presented. Finally, the conclusion is given in the last section.

\section{Review of FTO Models}

We start with a brief review of the basics of FTOs by introducing two typical models, the X-cube model~\cite{Checkerboard} and Hal\'{a}z-Hsieh-Balents (HHB) model~\cite{SpinChain}. After that we shall introduce the outstanding issue on the quantum phase transition of FTOs, as studied in this work.

\subsection{X-Cube model}
The X-Cube model is a spin-1/2 model defined on the edges of a cubic lattice~\cite{Checkerboard}, with the Hamiltonian
\begin{equation}
H_{\rm X-Cube} = -\frac{w}{4}\left(\sum_c A_c + \sum_{v,i} B_v^{(i)}\right),
\end{equation}
where $c$ runs through all cube centered at $c$ and $A_c$ is a product of 12 $\sigma_x$-operators defined on the edges of the cube, $v$ runs through all vertexes and $B_v^{(i)}$ is a product of 4 $\sigma_z$-operators defined on the edges which are attached to the vertex and are orthogonal to the direction $i$. $A_c$ and $B_v^{(i)}$ commute, so the ground state can be built from an eigenstate of all $B_v^{(i)}$. Moreover, since $A_c^2=B_v^{(i)2}=1$, the ground states can be written in the following form
\begin{equation}
\left|g\right> = \prod_c \frac{1+A_c}{\sqrt{2}}\left|\forall v,i, B_v^{(i)} = 1\right>.
\end{equation}
The forms of the ground states shows clearly that the topological GSD originates from different choices of $\left|\forall v,i, B_v^{(i)} = 1\right>$. The ground states thus generated can only be connected through flipping spins along straight lines that form noncontractible loops. In this model, the degeneracy is $2^{6N-3}$ for the system with $N^3$ cubes.

There are two types of excitations~\cite{Checkerboard}. (i) Lineons - generated by applying $\sigma^x$ on a ground state along an edge. The lineons are formed at the endpoints of the edge with energy $w/2$. If $\sigma^x$ is applied on consecutive edges that forms a straight line, two lineons will be formed on the head and on the tail. If the consecutive edges contain turning points, right at the turning points new excitations will be generated and cost extra energy. (ii) Fractons - generated by applying $\sigma^z$ on a ground state on the edges that forms a membrane perpendicular to the direction of the edges. The fractons then form on the corners of the membrane, and the energy of each group of four fractons is $w$.

\subsection{Hal\'{a}z-Hsieh-Balents (HHB) model}

The HHB model is a spin-1/2 model defined on a body-centered cubic (bcc) lattice~\cite{SpinChain}, where each lattice site contains four spins, and the Hamiltonian
\begin{equation}
H_{\rm HHB} = -J\sum_{\mathbf{r}j} \sigma^x_{\mathbf{r},j} \sigma^y_{\mathbf{r}+\mathbf{r}_j,j} - \lambda J \sum_{\mathbf{r}, j \neq j'} \sigma^z_{\mathbf{r}j} \sigma^z_{\mathbf{r} j'}.
\end{equation}
Here $\mathbf{r}$ runs over different lattice sites, $j = A,B,C$ or $D$ for each flavor of spin and $\mathbf{r}_A, \mathbf{r}_B, \mathbf{r}_C$ and $\mathbf{r}_D$ are the shortest vectors connecting two sites in the direction $\{111\}, \{\bar{1}11\}, \{1\bar{1}1\}$ and $\{11\bar{1}\}$ respectively. If $\lambda \rightarrow \infty$, all spins in the same site are fully polarized. For $\lambda \gg 1$, the first term of the Hamiltonian can be treated as a perturbation, and the effective Hamiltonian at the leading order can be written as
\begin{eqnarray}
H_{\rm HHB, eff} = -\frac{w}{32} \sum_\mathbf{r} \tilde{A}_\mathbf{r},
\end{eqnarray}
where $\tilde{A}_\mathbf{r} = \prod_{s=\pm j} \tilde{\sigma}^y_{\mathbf{r}+s\mathbf{r}_j}\prod_{s=\pm, j=\{B,C,D\}} \tilde{\sigma}^x_{\mathbf{r}+s(\mathbf{r}_A-\mathbf{r}_j)}$ and $w \sim J/\lambda^{31}$.
The tilded Pauli matrices at $\mathbf{r}$ denote the corresponding Pauli matrices defined on the subspace of two ground states in each site $\mathbf{r}$ at $\lambda = \infty$. All $\tilde{A}_\mathbf{r}$ commutes and they have eigenvalues $\pm 1$ so that in the similarity of the deviation in the X-Cube model, the ground states can be written in the following form
\begin{equation}
\left|g\right> = \prod_e \frac{1+\tilde{A}_e}{\sqrt{2}}\left|\forall o, \tilde{A}_o = 1\right>,
\end{equation}
where $e$ and $o$ are distinct subsets of the lattice sites such that each forms a simple cubic lattice that are translated in the $\{111\}$ direction.

Similarly, the topological GSD originates from different choices of $\left|\forall o, \tilde{A}_o = 1\right>$, and is $2^{12N-10}$ for $2N^3$ lattice sites. The system also hosts two different types of excitations. (i) Applying $\tilde\sigma_x$ on odd sites and $\tilde\sigma_y$ on even sites (or vice versa) along a straight line in the direction $\pm \mathbf{r}_j$ ($j=A,B,C,D$) yields a pair of lineon excitations. Unlike X-Cube model, the straight lines can be packed together to form 2D membrane or 3D body operators. (ii) Applying $\tilde\sigma_x$ on the sites along a line in $\pm(\mathbf{r}_A - \mathbf{r}_{j\neq A})$ directions also yields a pair of lineons. This operator can again be packed together to form 2D membrane or 3D body operators. Thus these types of excitations are classified as 3D fractons in our terminology.

\subsection{Overview and FTO breaking down}

All FTO models have topological GSD. The ground states can only be connected to each other through global operation, for example the spin-flipping along noncontractible loops. Therefore, any local perturbations cannot break the GSD. However, a strong enough global perturbation can break the GSD. For example, we consider a transverse magnetic term on X-cube model with
\begin{equation}\label{xcube2}
H = H_{\rm X-Cube} + v \sum_e \sigma_{x,e},
\end{equation}
where $e$ runs through all edges of the cubic lattice. This external term can generate lineons on the endpoints of the edges. It is easy to see that for $v\rightarrow\infty$, there is only one ground state with all spins being aligned along $-x$ direction, so a quantum phase transition must occur at some finite critical $v=v_c$, beyond which the FTO breaks down. This brings an outstanding question - How does the external term breaks the GSD? Answering the question is important to understand the topological phase region of FTO, and is crucial for the realization of FTO in real physical systems without fine tuning. Currently the FTOs are given in toy models which can be solved analytically. With an addition of external term, the system typically becomes not exactly solvable, and how to characterize the phase transition is a highly nontrivial but challenging issue.

\section{Quantum phase transition of FTO: definition and generic formalism}

\subsection{Define the quantum phase transition}

To approach this question, we consider the Hamiltonian in the following form
\begin{equation}
H = H_0 + \bar{v} V(v),
\end{equation}
where $H_0$ is unperturbed Hamiltonian of FTO with system size $N^3$, the external perturbation $V(v)$ is considered to the system with strength parameter $v$, and $\bar{v}$ is a dimensionless factor that will be taken as unity latter in the study. The perturbation may induce couplings in the ground states and thus break the GSD. Let $\Delta E(N,v)$ be the energy splitting of the ground state subspace. We shall see that,
\begin{equation}
\begin{cases}
\lim_{N\rightarrow\infty}\Delta E(N,v<v_c) = 0,\\
\lim_{N\rightarrow\infty}\Delta E(N,v>v_c) > 0.
\end{cases}
\end{equation}
This implies $v=v_c$ is a critical value, below which GSD is unchanged, while beyond which the GSD shall break down. In particular, we shall show that
when $v<v_c$, for any two degenerate ground states $|g\rangle$ and $|g'\rangle$, we have $\lim_{N\rightarrow\infty}\left<g\left|H_{\rm eff}\right|g'\right> = 0$, and for $v>v_c$, for any $|g\rangle$, there exists other ground states $|g'\rangle$ and $\lim_{N\rightarrow\infty}\left<g\left|H_{\rm eff}(v>v_c) \right|g'\right> \rightarrow\infty$. Here $H_{\rm eff}$ is an effective Hamiltonian defined on the subspace of the ground states that captures the transitions among them. Note that to avoid finite-size effect, the system size $N$ must be taken as infinity.

\begin{figure}[b]
\centerline{\includegraphics[width=1\columnwidth]{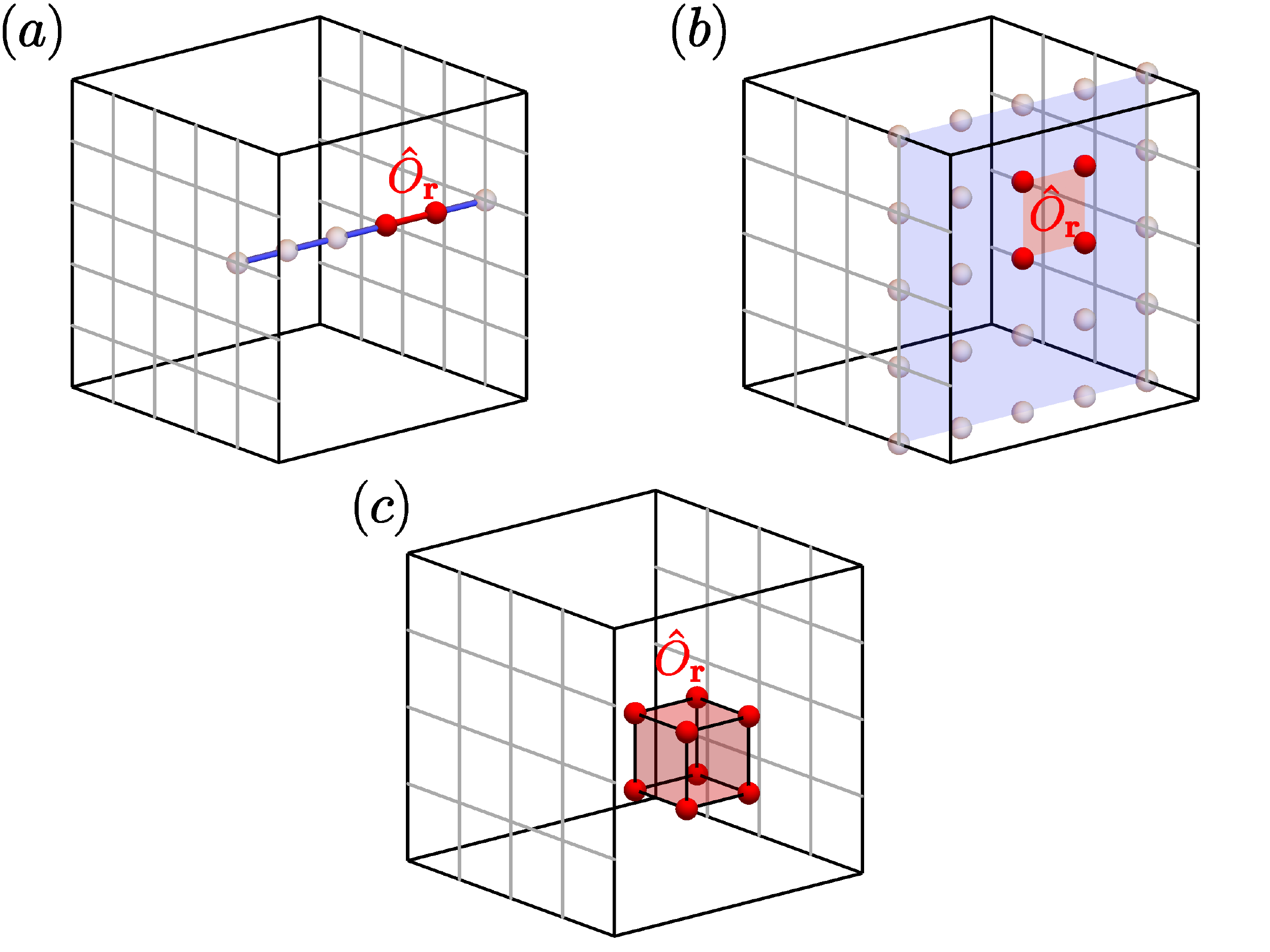}}
\caption{Schematic diagrams for the definition of $\hat{O}_\mathbf{r}$, when it produces (a) lineons, (b) 2D fractons or (c) 3D fractons in the vicinity. The red circles denote the lineons (2D fractons) produced by the operation of $\hat{O_\mathbf{r}}$. The operation of $\hat{O}$ can also annihilate the excitations for those points which excitation is already present. }
\label{DefO}
\end{figure}

\subsection{A general framework}
We consider that the external extra term is written in the following generic form of fracton operators
\begin{equation}
V = v\sum_\mathbf{r} \hat{O}_\mathbf{r},
\end{equation}
where $\hat{O}_\mathbf{r}$ generates or annihilates two lineons (Fig. \ref{DefO}(a)), four 2D fractons (Fig. \ref{DefO}(b)) or eight 3D fractons (Fig. \ref{DefO}(c)) at its vicinity. In Fig.~\ref{DefO}, the operator $\hat{O}_\mathbf{r}$ is denoted by red line, square or cube, respectively, and the excitations created by it are represented by the red spheres.

To facilitate the study on how the external term $V(v)$ changes GSD, we split the Hilbert space into two subspaces
\begin{equation}
\psi = \left(\begin{matrix}
\psi_g\\
\psi_{\bar{g}}
\end{matrix}\right),
\end{equation}
where $\psi$ represents the full Hilbert space, $\psi_g$ denotes the ground state subspace, and $\psi_{\bar{g}}$ denotes the subspace of the states with excitations. With this notation, we project the whole Hamiltonian onto the ground state subspace, and obtain the effective Hamiltonian (Appendix)
\begin{eqnarray}\label{effectiveapproach}
H_{\rm eff} &=& H_{0,gg}+\bar{v} V_{gg} - \sum_{i=2}^{\infty} \bar{v}^i V_{g\bar{g}}\times\nonumber\\
&&\left(\sum_{(s_1,\cdots,s_m)\in P_{i-1}} Q^{(s_1)}V_{\bar{g}\bar{g}}\cdots V_{\bar{g}\bar{g}}Q^{(s_m)}\right)V_{\bar{g}g}. \hspace*{15pt}
\end{eqnarray}
$P_{i-1}$ is the set of integer partitions of $i-1$, i.e. $s_1+s_2+\cdots+s_m = i-1$. The subscripts $gg$ (or $\bar{g}\bar{g}$) means that the corresponding operator is projected onto the ground state subspace (or the excited state subspace). Similarly, the subscripts $g\bar{g}$ and $\bar{g}g$ mean that the corresponding operators connect the ground state with excited state subspace. Finally, $Q^{(s)}$'s are the factors of inverse energy scale, which denotes the energy difference between ground state and intermediate excited states of the FTO (see more details in the Appendix). Since only the operation on noncontractible loops can induce transition between two different ground states, the leading order correction to the ground state subspace must be the $N$-th order. In the leading order contribution, only the terms with $Q^{(1)} = (E_0 - H_{0,\bar{g}\bar{g}})^{-1}$ have non-zero matrix element between two different ground states. Therefore, the effective Hamiltonian by taking into account the leading order contribution reads
\begin{equation}
H_{\rm eff}^{(N)} = -\bar{v}^N V_{g\bar{g}} \left(Q^{(1)} V_{\bar{g}\bar{g}} \cdots V_{\bar{g}\bar{g}}Q^{(1)}\right) V_{\bar{g}g}.
\label{GSmatrixelementP}
\end{equation}
Any lower order perturbation cannot couple different ground states, since a noncontractible loop must be formed by at least $N$ times of $\hat{O}_\mathbf{r}$ operations.

The energy splitting in the ground states depends on two aspects. The first is the coupling between each two ground states (e.g. $|g\rangle$ and $|g'\rangle$) up to all higher orders. 
In particular, the matrix element for the ($N+r$)th-order contribution is given by $\mathcal{H}^{(N+r)}_{{\rm eff},gg'}=\langle g|H_{\rm eff}^{(N+r)}|g'\rangle$, with $r=0$ being the leading-order contribution. The second is the number of ground states $\left|g'\right>$ which are coupled to a certain fixed ground state $\left|g\right>$ by the external term, denoted by $f^{(N+r)}$. The energy splitting from the asymptotic behavior should be given by
\begin{eqnarray}\label{splitting1}
\Delta E \sim \sum_{r=0}^\infty f^{(N+r)} \mathcal{H}_{{\rm eff},gg'}^{(N+r)}, \ N\rightarrow\infty.
\end{eqnarray}
Intuitively, the ground state subspace can be treated as a flat band of strongly correlated many-body states before phase transition. For the strength of external perturbation exceeding a critical value, the flat band is deformed to be a dispersive one, with $\Delta E$ being the bandwidth.

\begin{figure}[t]
\centerline{\includegraphics[width=\columnwidth]{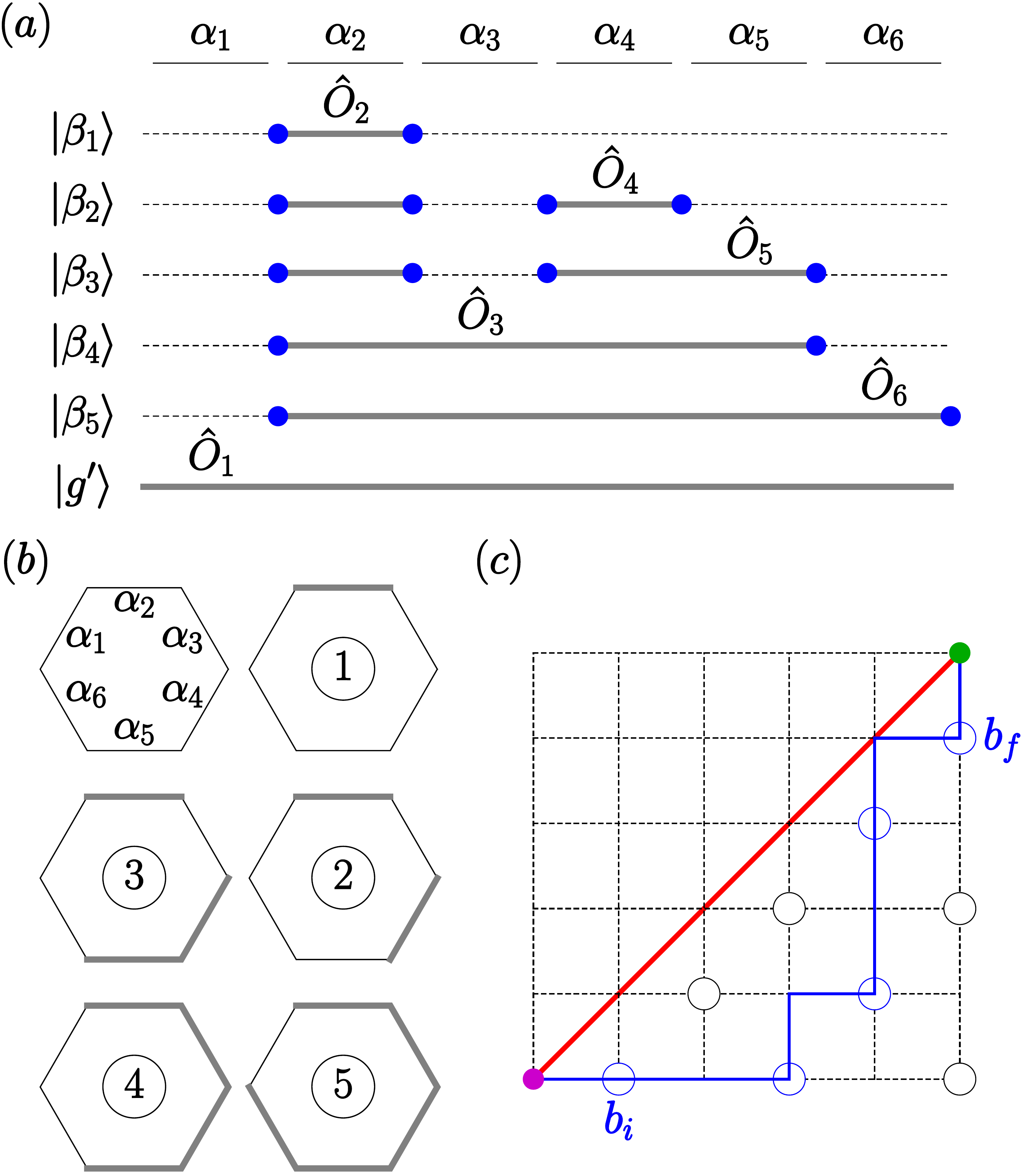}}
\caption{(a,b) One possible transition described in $H^{(N=6)}_{\rm eff}$ from $\left|g\right>$ to $\left|g'\right>$. $\hat{O_2}$, $\hat{O_4}$, $\hat{O_5}$, $\hat{O_3}$, $\hat{O_6}$ and $\hat{O_1}$ are applied on the ground state $\left|g\right>$ successively to produce $\left|\beta_{1\sim 5}\right>$ and $\left|g'\right>$. $\alpha_i$ are lattices that $\hat{O}_\mathbf{r}$ defined on and lineons are defined in between two neighboring lattices and are shown as blue dots. (c) Schematic diagram showing the definition of $5^{\rm th}$ Catalan's number $\mathsf{C}_5$.}
\label{IntMedState}
\end{figure}
\section{Critical point of the quantum phase transition}

We proceed to study the key issue, the critical point of the quantum phase transition in the FTOs. We shall show that, for the external perturbation creating only lineon type excitations, the critical point of the phase transition can be given by a universal formula and in some special cases the result given by the formula is exact, as presented in the following theorem. 

{\em {\bf\em Theorem 1}: Let $w$ be the energy of a pair of separated lineons and $w_i$ be the energy of two lineons along different directions crossing at an intersecting site. Then the quantum phase transition occur exactly at $v=v_{c1}=\frac{w}{4}$ if $w_i\geq w$ or $v=v_{c1}'\approx \frac{w}{4} \times (1-\frac{w-w_i}{w}\times 0.26)$ if $w_i<w$.} This theorem can be proved by solving the leading-order and all higher-order contributions in the thermodynamical limit.    We first begin our discussion on the general properties of the contributions of different order in the subsections A and B. In the latter part of subsection B and subsection C, we assume that $f=w$ and prove the theorem in this regime. In subsection D, we argue that the exact result can extend to the case $f>w$ and employ a similar method to give a universal formula for $f\leq w$ under certain approximation.

\subsection{The leading order contribution}
\begin{figure}[t]
\centerline{\includegraphics[width=1\columnwidth]{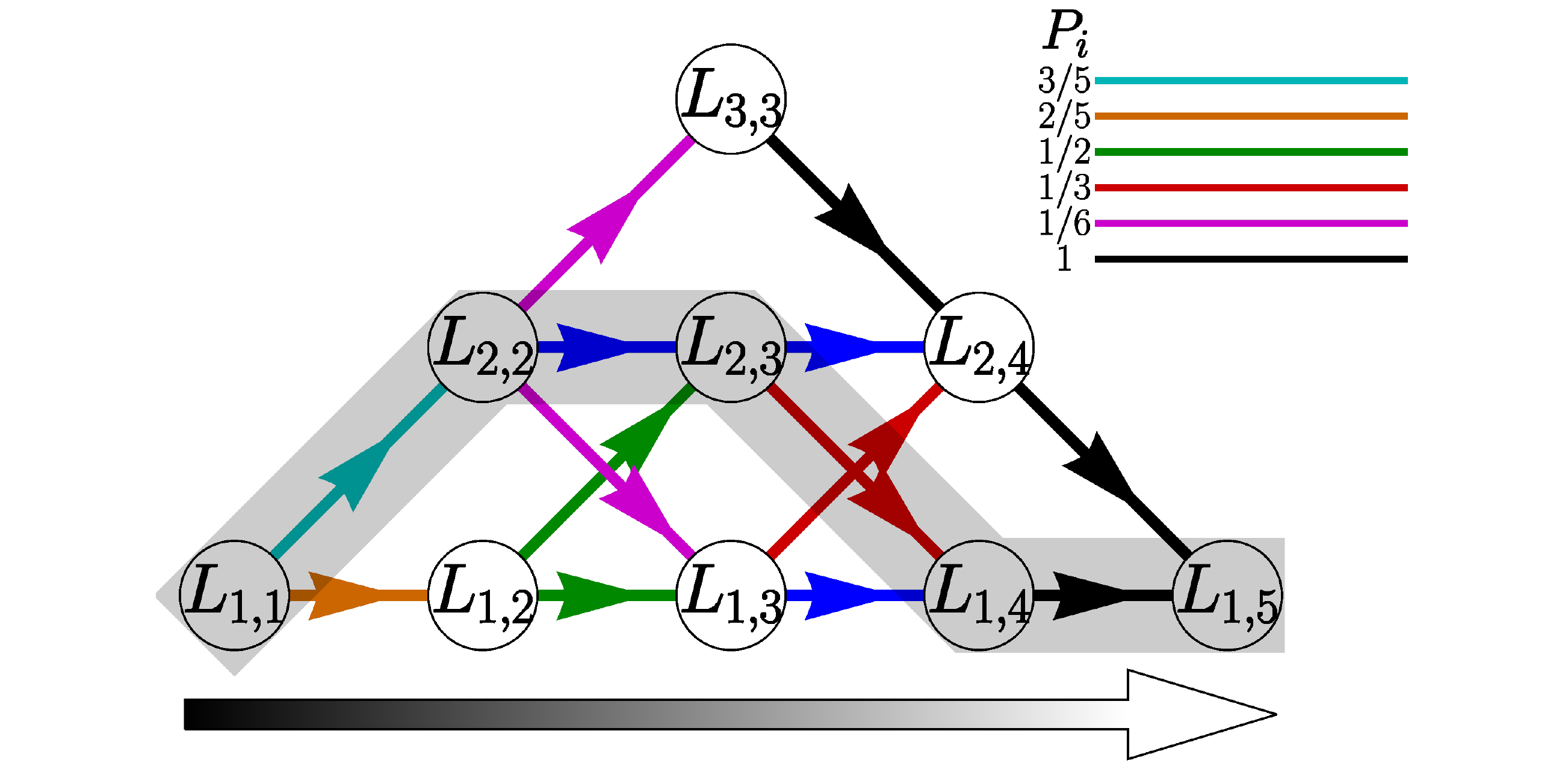}}
\caption{An equivalent random walk network for $N=6$. The shaded path depicts the sum of all possible intermediate states so that $F_{\beta_{1\sim 5}}$ are 1,2,2,1 and 1 respectively, which Fig. \ref{IntMedState}(a,b) depict one example of such paths.}
\label{RandomWalkL}
\end{figure}

We start with the leading order contribution to the matrix element of $H_{\rm eff}$ between two ground states $\left|g\right>$ and $\left|g'\right>$. Let the two ground states be connected by a noncontractible straight line of $\hat{O}_\mathbf{r}$ operators, i.e.
\begin{equation}
\left|g'\right> = \prod_r \hat{O}_\mathbf{r} \left|g\right>,
\end{equation}
where $\mathbf{r}$ runs through all sites along a straight line. Denote by $V_{\gamma\gamma'} = \left<\gamma\left|V\right|\gamma'\right>$ the matrix element of $V(v)$ and $F_{\gamma}$ the number of pairs of generated lineons at the intermediate states $\left|\gamma\right>$. The matrix element of the effective Hamiltonian reads
\begin{eqnarray}\label{leadingorder}
\mathcal{H}_{{\rm eff},gg'}^{(N)} &=& \left(\frac{-1}{w}\right)^{N-1}\sum_{\{\beta_1,\cdots, \beta_{N-1}\}} \frac{V_{g\beta_1}V_{\beta_1\beta_2}\cdots V_{\beta_{N-1} g'}}{F_{\beta_1}F_{\beta_2}\cdots F_{\beta_{N-1}}},\nonumber\\
&=& v \left(\frac{-v}{w}\right)^{N-1} {\sum_{\{\beta_1,\cdots, \beta_{N-1}\}}}' \frac{1}{F_{\beta_1}F_{\beta_2}\cdots F_{\beta_{N-1}}}.
\end{eqnarray}
The denominator implies that the energy of the intermediate state depends purely on the number of lineons in the system. 
The summation is over all permutation of $\{\beta_1,\cdots, \beta_{N-1}\}$, the last line in above equation is obtained by noticing that $V_{\gamma\gamma'}$ is either $0$ or $v$, and the prime means that the summation is over all the intermediate states with $V_{\gamma\gamma'}=v$. As an example, when $N=6$, one of the terms in the summation is shown in Fig. \ref{IntMedState}(a,b), where the number of lineons $F_{\beta_{1\sim 5}}$ of the intermediates states $\left|\beta_{1\sim 5}\right>$ are $1,2,2,1$ and $1$ respectively.

Exactly counting the summation in Eq.~\eqref{leadingorder} is highly nontrivial when $N$ is large. With the details presented in the Supplementary Material~\cite{supplementary}, we find that the exact value of the leading order perturbation takes a novel form
\begin{eqnarray}\label{leadingorderresult}
\mathcal{H}_{{\rm eff},gg'}^{(N)} = v\bigr(-\frac{v}{w}\bigr)^{N-1}C_{2N-2}^{N-1}\simeq \frac{v}{\sqrt{\pi N}}\bigr(-\frac{4v}{w}\bigr)^{N-1}.
\end{eqnarray}
The leading order contribution is deeply related to the so-called Catalan number~\cite{supplementary}, $C_{2N-2}^{N-1}/N$, as illustrated in Fig.~\ref{IntMedState}(c). The $N^{\rm th}$ Catalan number is defined as the number of paths along the edges of a grid with $N \times N$ square cells from the bottom-left corner (purple dot) to the top-right corner (green dot) with two restrictions. (i) Each step of the path has to be either rightward or upward. (ii) The whole path has to be below the diagonal line (painted red). Or equivalently, the path has to pass through $N$ odd lattices below the diagonal, including the lattice rightward to the bottom-left corner ($b_i$), the lattice downward to the top-right corner ($b_f$) and $N-2$ more odd lattices (circles on the vertexes of the grid). In particular, the $N-1^{\rm th}$ Catalan's number ${\cal C}_{N-1}$ is related to the leading order perturbation by $\mathcal{H}_{\rm eff,gg'}^{(N)}\propto N {\cal C}_{N-1}$.

To better understand the leading order contribution, we show below that it can be mapped to a random walk problem, which has the important merit to facilitate the further study of higher-order contributions and the more generic extra terms creating 2D/3D fractons. More specifically, we consider a random walk starting at $L_{1,1}$ in a network $L_{i,j}$, where $1\leq i \leq \text{min}(j,N-j)$ equals the number of pairs of lineons in the $j$-th step of the intermediate process, with $1\leq j \leq N-1$, and moving $N-2$ steps with probabilities in each step~\cite{supplementary}
\begin{eqnarray}
P|_{L_{i,j}\rightarrow L_{i-1,j+1}} &=& \frac{i(i-1)}{(N-j)(N-j-1)}, \nonumber\\
P|_{L_{i,j}\rightarrow L_{i,j+1}} &=& \frac{2i(N-i-j)}{(N-j)(N-j-1)}, \nonumber \\
P|_{L_{i,j}\rightarrow L_{i+1,j+1}} &=& \frac{(N-i-j)(N-i-j-1)}{(N-j)(N-j-1)}.\nonumber
\end{eqnarray}
For each step, we set the reward gained by the random walk as $1/i$ so that the mean of the product of rewards matches the matrix element $\mathcal{H}_{{\rm eff},gg'}^{(N)}$ after summing over all paths, except for the prefactor $v\left(-v/w\right)^{N-1}$. The exact result is given by
\begin{eqnarray}
\mathcal{H}_{{\rm eff},gg'}^{(N)} = \sum_{i_2=1}^{2} \sum_{i_3=1}^3\cdots \sum_{i_{N-2}=1}^2 \frac{\prod_{j=1}^{N-2} P|_{L_{i_j,j}\rightarrow L_{i_{j+1},j+1}}}{\prod_{j=1}^{N-1}i_j}
\end{eqnarray}
with $i_1=i_{N-1}=1$.
A simple example with $N=6$ of the random walk network is shown in Fig. \ref{RandomWalkL}. The shaded path shows a case where $i_2=i_3=2$ and $i_4=1$ and it corresponds to the sum of all possible intermediate states so that $F_{\beta_{1 \sim 5}}$ are $1,2,2,1$ and $1$ respectively. One example of such states are shown in Fig. \ref{IntMedState}(a). The sum of the rewards of the random walk equals the $N$ times of the Catalan number, recovering the result in Eq.~\eqref{leadingorderresult}.

\subsection{Higher order contributions}\label{HOC}

The mapping to random walk problem can be generalized to the study of all higher-order contributions $\mathcal{H}_{{\rm eff},gg'}^{(N+r)}$, with $r$ being even integers. Note that the odd $r$ is forbidden since the effective Hamiltonian with an addition of odd number of $\hat{O}_{\bold r}$ operations must scatter ground states to excited ones. It is convenient to consider the higher order terms in the two different case, namely, the higher-order contribution with finite $r$ and the contribution with infinite $r$, respectively.

\begin{figure}[t]
\centerline{\includegraphics[width=1\columnwidth]{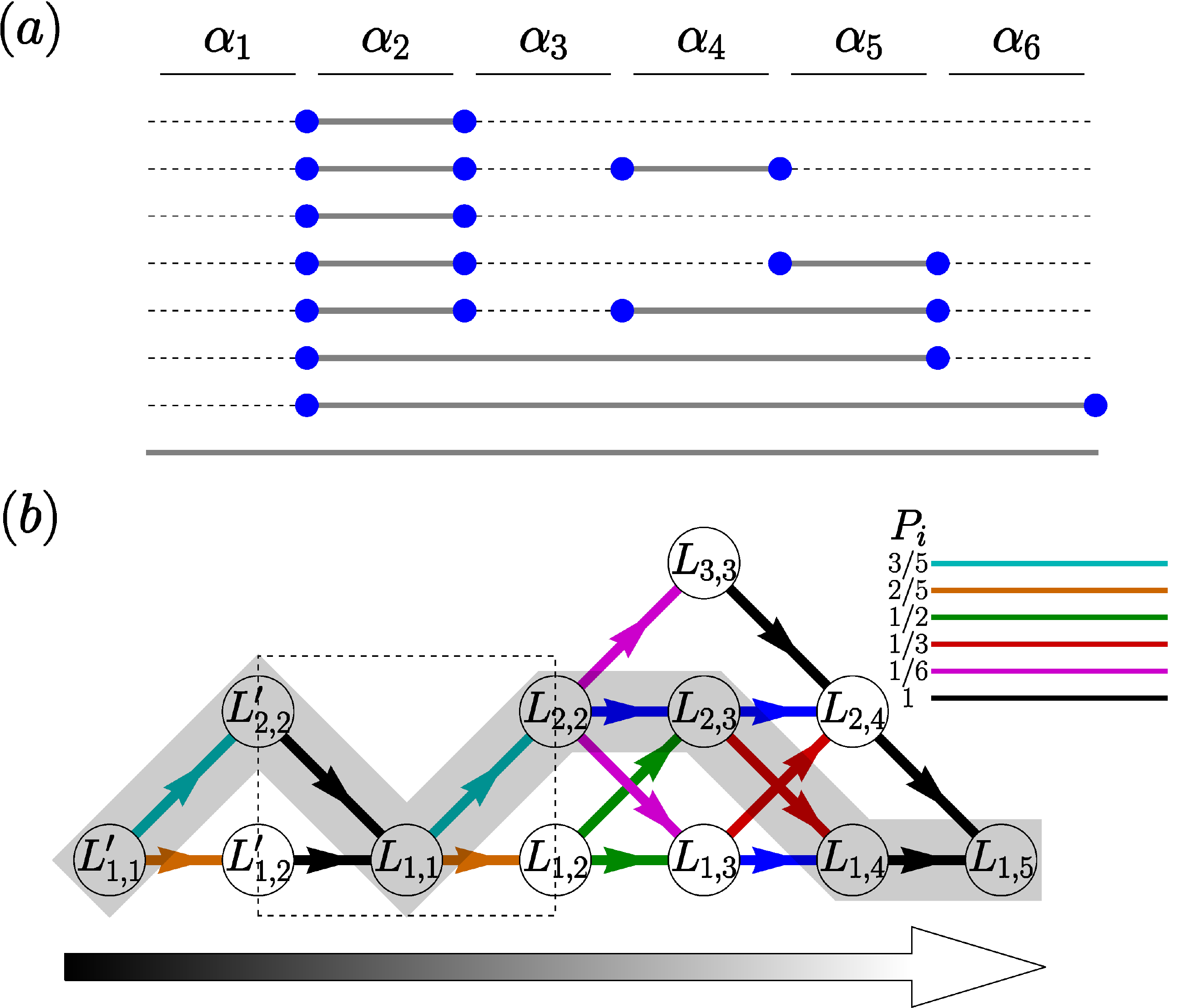}}
\caption{(a) One possible way to transit from $\left| g \right>$ to $\left| g'\right>$ described by $H^{(N+r)}_{\rm eff}$, where $N=6$ and $r=2$. (b) A transformed network $L^{(1)}$ so that the random walker move back at the second step. The shaded path depicts the sum of all possible intermediate states so that $F_{\beta_{1\sim 7}}$ are $1,2,1,2,2,1$ and $1$ respectively, which (a) depict one example of such paths. The dashed box indicates the inserted part of the new network $L^{(1)}$ to account for the step that moves backward in the original network $L$.}
\label{RandomWalkHFig}
\end{figure}

To get the intuition of the higher-order contributions, we consider first the lowest higher order term with $r=2$. This type of higher order terms contain $N+2$ number of $\hat{O}_\mathbf{r}$ operators and some of them has to be repeated. Since $\hat{O}_\mathbf{r}^2=1$, the repetition is actually equivalent to a backward step in the random walk process. For $r=2$, the network $L$ is allowed to move back by one step at an intermediate step. As a simple example with $N=6$ and $r=2$ shown in Fig.~\ref{RandomWalkHFig}(a), the $\hat{O}_4$ operation is repeated at the third step. When represented by the random walk, this process is equivalent to moving back at the second step. Specifically, the network $L$ is expanded to a new one $L^{(1)}$, in which the random walk includes a new boxed sub-network to account for the process of moving backward by two steps, as shown in Fig. \ref{RandomWalkHFig}(b). 
For the generic configuration, let $P_i$ be the probability that the walker moves back at $i$-th step. The average reward, and so the matrix element, can be approximated as
\begin{equation}
\mathcal{H}_{{\rm eff},gg'}^{(N+2)} \approx \sum_{i=3}^{N} P_i R_{N,i}.
\end{equation}
Here $R_{N,i}$ is the average reward of the random walk process which includes the backward moving at the $i$-th step. It can be shown that whenever $P_i$ is finite, the asymptotic behavior of $R_{N,i}$ remains the same as $\mathcal{H}_{{\rm eff},gg'}^{(N)}$ for large $N$ (see Supplementary Material~\cite{supplementary}). This further determines the asymptotic behavior of $\mathcal{H}_{{\rm eff},gg'}^{(N+2)}$. For generic finite $r$, the above argument iterates $r/2$ times to form a new network $L^{(r/2)}$ and the similar result can be obtained. It then follows that that the matrix element $\mathcal{H}_{{\rm eff},gg'}^{(N+r)}$ with finite $r$ should be dominated by $(4v/w)^{N+r}$ and the sum of the perturbation series up to such higher orders is exponentially small if $v<w/4$.

The remaining question is how the perturbation grows as $r$ tends to infinity, noting that the iteration cannot be done by infinite number of times. For this purpose, we first assume that the energy of 2 lineons coming from mutually different directions meeting at the same point is $2 \times w/2 = w$, i.e. lineons coming from different directions are independent. Let $r=\zeta N$, where $\zeta$ is a real number. After some algebra we find for higher order perturbations with infinite $r$ that~\cite{supplementary}
\begin{equation}
\lim_{N \rightarrow \infty} \frac{\mathcal{H}_{{\rm eff},gg'}^{(N+\zeta N+1)}}{\mathcal{H}_{{\rm eff},gg'}^{(N+\zeta N)}} = \frac{\bar{\lambda}}{N} \times \frac{v}{w},
\end{equation}
where $\bar{\lambda} = N/4$ is the reward average over all possible configurations, i.e. all possible intermediate states connecting $\left|g\right>$ with $\left|g'\right>$. The above is an important novel result, and can be understood in the following way. For the higher order term with infinite $r$, the network $L^{(r/2)}$ allows to move backward infinite number of times so that the process are close to unrestrained random walk, namely, the random walk can perform in arbitrary directions. Thus the growth of the average of product of rewards is captured by the reward averaging over all possible configurations. With this one can show that the matrix element grows by $4v/w$ as $r$ increases one order.~\cite{supplementary}.

\subsection{Energy splitting}
\begin{figure}[t]
\centerline{\includegraphics[width=\columnwidth]{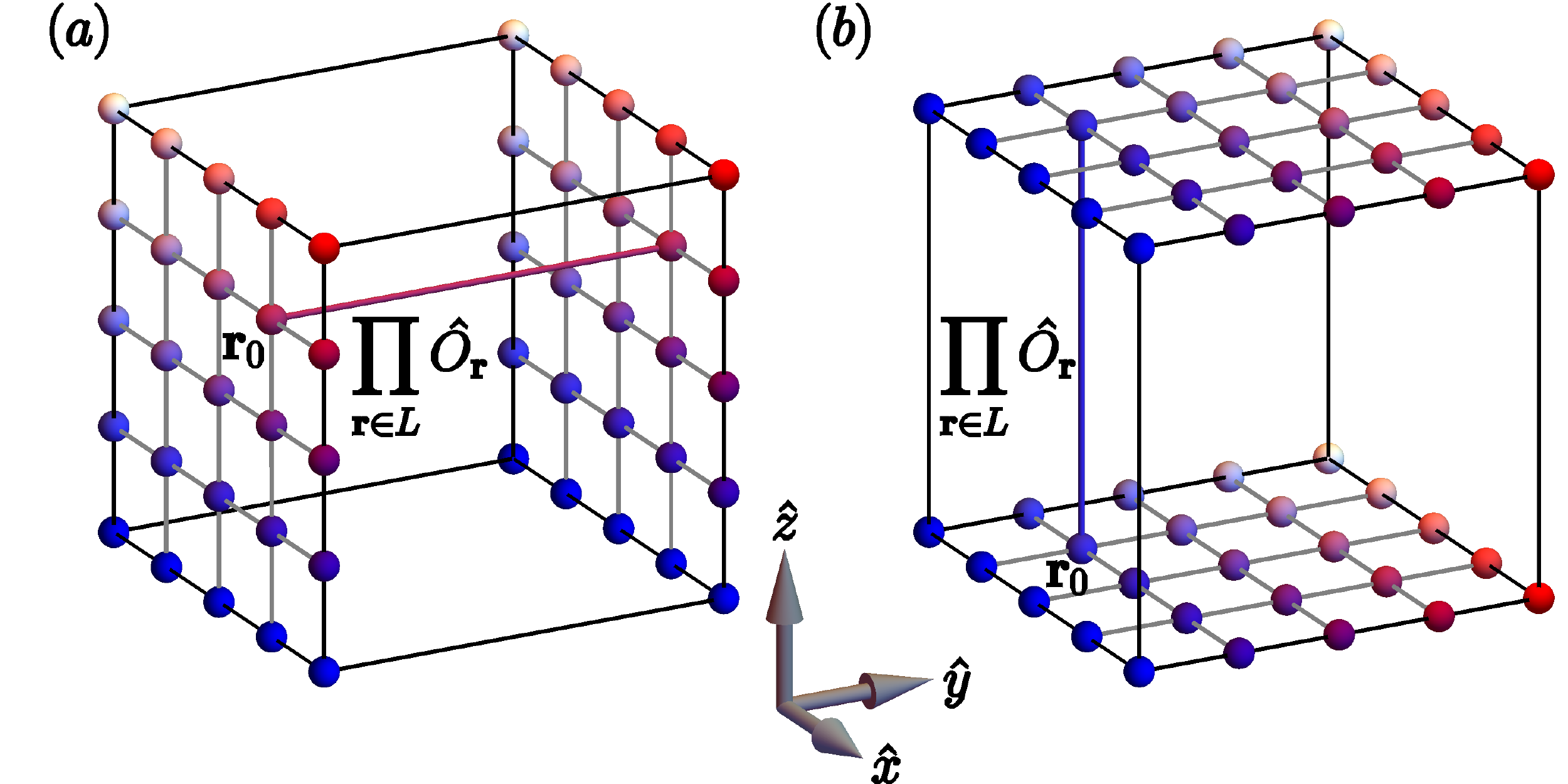}}
\caption{Schematic diagrams showing the order of magnitude of $f^{(N)}$. The lines in the diagrams represent the product of operators $\hat{O}_\mathbf{r}$ along a straight line $L$ to transform a ground state to another. Each choice of $\mathbf{r}_0$ represents one possible transformation and so each choice contributes $1$ to $f^{(N)}$ so that pairs of surfaces give contributions of $N^2$ to $f^{(N)}$. (a) and (b) show two examples that the pairs of surfaces are oriented in the $\hat{y}$ and $\hat{z}$ directions, respectively. This means that the final $f^{(N)}$ is of the order of $N^2$.}
\label{EnergySplit}
\end{figure}
Summing over all-order contributions yields the total transition matrix element
\begin{equation}\label{Heff}
\mathcal{H}_{{\rm eff},gg'} = \sum_r \mathcal{H}_{{\rm eff},gg'}^{(N+r)} = \sum_r \mathfrak{C}^{(r)} \left(\frac{4v}{w}\right)^{(N+r)}.
\end{equation}
As the growth of $\mathcal{H}_{{\rm eff},gg'}^{(N+r)}$ is $4v/w$ as $r\rightarrow \infty$, the growth of prefactor $\mathfrak{C}^{(r)}$ is subexponential. On the other hand, the number $f^{(N)}$ obeys power-law of the system size and is of the order $N^2$. This is because for a reference ground state $\left|g\right>$, the number of other ground states that can be coupled to it through $N$ $\hat{O}_\mathbf{r}$ operators is $\sim N^2$ (see Fig.~\ref{EnergySplit}). Further, the number $f^{(N+r)}$ does not grow until $r$ is comparable to $N$, since an addition of finite number of $\hat{O}_\mathbf{r}$ operations couples no new ground states. Thus the number $f^{(N+r)}$ also grows subexponentially versus $N$ and $r$. With these results we conclude from Eq.~\eqref{splitting1} that the system, which we assume the lineons coming from different directions are independent, has a critical point given by
\begin{eqnarray}
v_{c1}=\frac{w}{4}.
\end{eqnarray}
For the perturbative term weaker than the above value the GSD of the FTO is unaffected. 

We provide an intuitive understanding of the above critical point by partially relating the perturbative study to the 1D transverse Ising model. Consider the case for two ground states $\left|g\right>$ and $\left|g'\right>$ which can be connected through applying $\hat{O}_\mathbf{r}$ operators on a single straight line, i.e. $\left|g\right> = \prod_{r\in L} \hat{O}_\mathbf{r} \left|g'\right>$. Here $L$ is a straight line forming a non-contractible loop. We further consider the following Hamiltonian $\tilde H=\tilde H_0+\tilde H_1$:
\begin{eqnarray}
\tilde H_0 = \frac{w}{2} \sum_i \hat{\xi}_{i,i+1}, \ \ \tilde H_1 = v \sum_i \hat{O}_i,
\end{eqnarray}
where $\hat{\xi}_{i,i+1}$ is an operator with eigenvalue $1$ when there is a lineon between site $i$ and $i+1$, and $0$ when there is not. $\hat{O}_i$ remove (create) when there is (not) a lineon in the vicinity of the site $i$. It then follows that
\begin{equation}\label{commutation}
\left[\hat{O}_i, \xi_{j,j+1}\right] = \hat{O}_i \left(1-2\hat{\xi}_{j,j+1}\right) \left(\delta_{ij}+\delta_{i,j+1}\right).
\end{equation}
With the above results we can verify that the matrix element of $\tilde H$ between $|g\rangle$ and $|g'\rangle$ is identical to $\mathcal{H}_{{\rm eff},gg'}$. Consider now the mapping that $\hat{\xi}_{i,i+1} \rightarrow (1+\sigma^z_i \sigma^z_{i+1})/2$ and $\hat{O}_i \rightarrow \sigma^x_i$,
which keeps the above commutation relation~\eqref{commutation}. This leads to $\tilde H\rightarrow H_{\rm Ising} = \frac{w}{4} \sum_i \sigma^z_i \sigma^z_{i+1} +  v \sum_i \sigma^x_i$, which is the 1D transverse Ising model with the critical point given by $v_c = w/4$. From this result we can see that at the critical point the bulk gap of the FTO transition should close. It is noteworthy that this mapping is valid for part of the configurations, not exactly identical to the total 3D effective Hamiltonian~\eqref{effectiveapproach}, but gives the correct critical point if the lineons coming from different directions are independent. Also the mapping is valid only for the regime before the phase transition.

\subsection{Correction due to intersecting lineons}

More generally, the result of critical value for system with lineon excitation depends on the energy $w_i$ of two lineons along different directions crossing at intersecting site. (i) $w_i = w$. The global line operators can be treated as independent so the critical point is $v=v_{c1}=w/4$. (ii) $w_i > w$. Then the matrix elements between two arbitrary ground states may be finite only at some value $v > w/4$. However, the critical point remains at $v=v_{c1}=w/4$ since the leading order contribution between $\left|g\right>$ and $\left|g'\right>$ that are connected by a single global line operator is finite at $v=v_{c1}=w/4$. (iii) $w_i < w$. the above calculated critical point is overestimated since the energies of the intermediate steps are overestimated. It can be seen from Eq. (\ref{leadingorder}) or the corresponding formula for higher order contributions that matrix elements of the effective Hamiltonian are proportional to the inverse of the product of energies of intermediate states. Therefore, the critical point depends on the average correction to the intermediate energies. To capture the effect analytically, a factor $\mathfrak{f}^{(r)}$ is inserted into the expression (\ref{Heff}) so that $\mathcal{H'}_{\text{eff},gg'} = \sum_r \mathfrak{C}^{(r)} \left(4v\mathfrak{f}^{(r)}/w\right)^{N+r}$. An estimation of the correction factor can be given below (details are found in the Supplementary Material~\cite{supplementary}.

We start the estimation by noting that the correction factor can be written as
\begin{equation}
\left(\mathfrak{f}^{(r)}\right)^{N+r} = \left<\prod_i \mathfrak{f}_{\beta,i}^{(r)}\right>_\beta = \frac{\sum_\beta \prod_i^{N+r}F_{\beta_i}^{-1}\mathfrak{f}^{(r)}_{\beta,i}}{\sum_\beta \prod_i^{N+r}F_{\beta_i}^{-1}},
\end{equation}
where $i$ denotes the $i^{\rm th}$  intermediate step, $\left<\cdots\right>_\beta$ denotes the weighted average over all configurations $\beta$ and $\mathfrak{f}_{\beta,i}$ captures the ratio of energy changes due to intersecting lineons. For example, if there are 4 non-intersecting lineons and 2 intersecting in a given configuration and the energy of each intersecting lineon is $w/3$, then $\mathfrak{f}_{\beta_i} = (4\times w/2+2\times w/3)/(6\times w/2) = 9/8$. We start the estimation of $\mathfrak{f}^{(r)}$ by assuming $\mathfrak{f}_{\beta,i}$ to depend on $F_{\beta_i}$ only. This assumption is reasonable since the probability of lineons to be intersecting is proportional to the density of lineons. With this assumption, the correction factor becomes
\begin{equation} \label{correctionfapprox}
\left(\mathfrak{f}^{(r)}\right)^{N+r}\approx \frac{\sum_{F_{\beta_1}F_{\beta_2}\cdots} \prod_i^{N+r} M_{\beta_i} F_{\beta_i}^{-1} \mathfrak{f}^{(r)}_{F_{\beta_i}}}{\sum_{F_{\beta_1}F_{\beta_2}\cdots} \prod_i^{N+r} M_{\beta_i} F_{\beta_i}^{-1}},
\end{equation}
where $M_{\beta_i}$ is the number of configurations for a given $F_{\beta_i}$. The correction factor $\mathfrak{f}_{F_{\beta_i}}^{(r)}$ can be calculated for each $F_{\beta_i}$ and the value of $\mathfrak{f}_{F_{\beta_i}}^{(r)}$ where $M_{\beta_i} F_{\beta_i}^{-1}$ maximizes is~\cite{supplementary}
\begin{equation}
\mathfrak{f}^{-1}_{\text{dom}} \approx 1 - \frac{w-w_i}{w}\times 0.26,
\end{equation}
where $w_i\leq w$ is the energy of two intersecting lineons and the critical point of this more generic case can be approximated by 
\begin{equation}
v_{c1}' = v_{c1} \times \mathfrak{f}^{-1}_{\text{dom}}.
\end{equation}
This result is universal for the type of extra terms that create only lineon excitations. In particular, the quantum phase transition of the transverse field X-cube model, as given in Eq.~\eqref{xcube2}, can directly obtained from the above result, and $v_\text{X-Cube}\approx (0.87 \pm 0.04)\times w/4$. The origins of the error are discussed in the supplementary material \cite{supplementary}. The value agrees with the numerical result $v_\text{X-Cube}^\text{pCUT}\approx 0.9196\times w/4$ by the method of perturbative continuous unitary transformations \cite{FractonRobustness}.
\section{Critical points for the generic extra terms}

The random walk theory developed based on the effective Hamiltonian approach can be further applied to study the more generic cases beyond lineon type perturbations. For those cases, the external perturbation can induce also the 2D and/or 3D fractons, e.g. for transverse field coupling to $\tilde\sigma_{x,y}$ in the HHB model,
namely, the term $\hat{O}_\mathbf{r}$ generates or annihilates four 2D fractons (as shown in Fig. \ref{DefO}(b)) or eight 3D fractons (as shown in Fig. \ref{DefO}(c)) nearby the site $\bold r$. We expect that when the strength $v$ of the extra term $V(v)$ exceeds certain critical value, the phase transition occurs and the FTO breaks down. Similar to the study for lineon type perturbations, the exact critical points can be obtained in three key steps. (i) Identify how the leading order contribution grows as $N\rightarrow \infty$. This encodes also how the higher-order contributions grow with finite $r$. (ii) With fixed $N$, identify how the infinite-$r$ higher-order contributions grow. (iii) Summing over all order contributions to obtain the energy splitting. 
Similarly, for a fixed ground state $\left|g\right>$, the number $f^{(N+r)}$ of $\left|g'\right>$ coupled to it through leading-order perturbation must be power-law function in $N$. This value does not grow until $r$ is comparable to or over $N$ (for 2D and 3D fractons). Thus for $N\rightarrow \infty$ the growth of $f^{(N+r)}$ is again subexponential.

{\em Upper limits.}--The exact result of the critical point is not analytically available in the current stage. The challenge arises from working out the leading-order contribution, as studied later. However, the higher-order contributions with infinite $r$ can be exactly obtained by the random walk theory. 
For the case with 2D or 3D fractons, the sole consideration of (ii) gives an upper limit $v_{c}^{(u)}$ of the critical point, 
as stated below.

\begin{figure}[t]
\centerline{\includegraphics[width=\columnwidth]{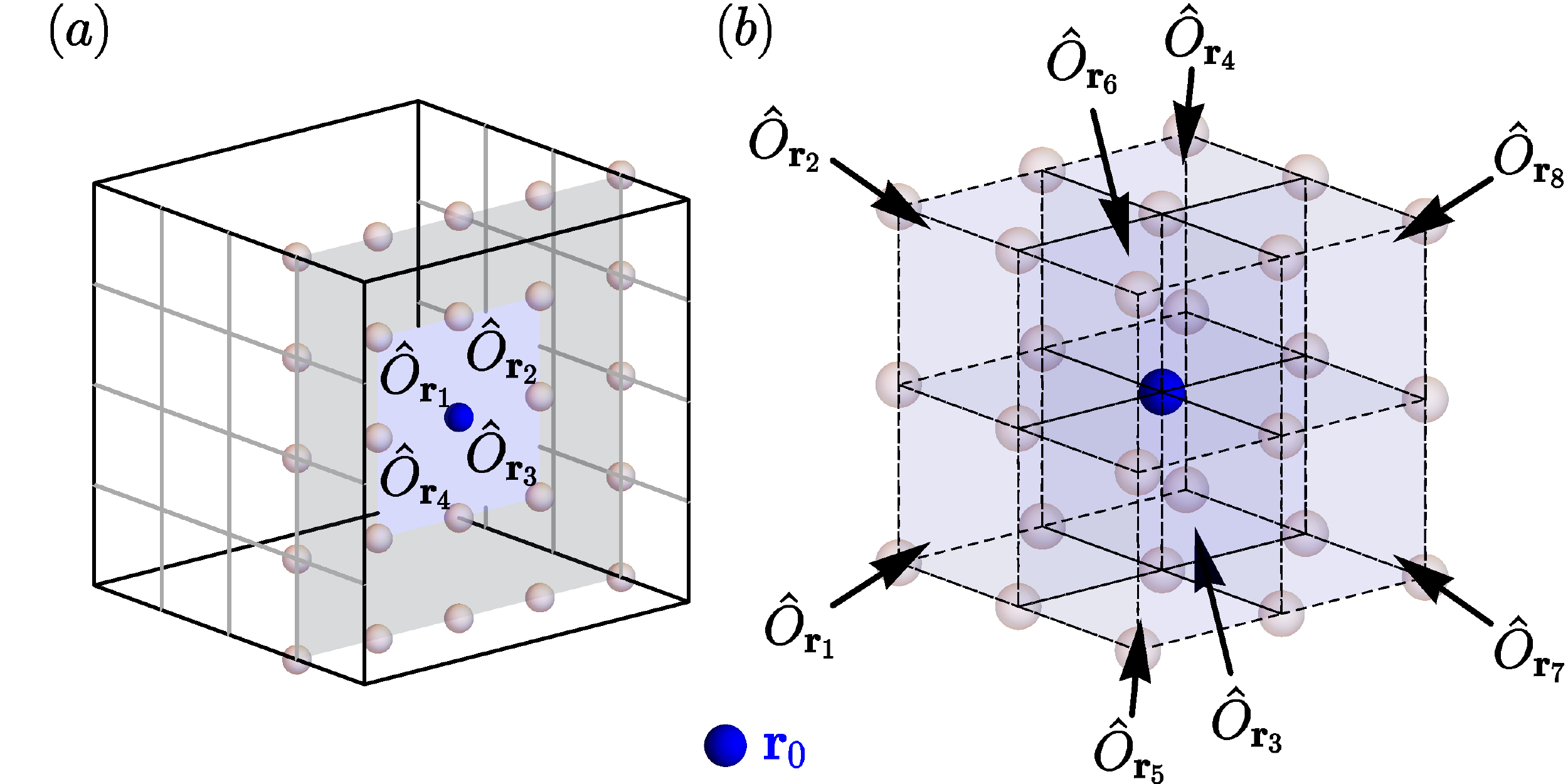}}
\caption{Schematic diagrams showing the definition of the operators $\hat{O}_{\mathbf{r}_i}$ surrounding the point $\mathbf{r}_0$ (blue sphere) on which a (a) 2D or (b) 3D fracton is defined. The (a) membrane or (b) body operators $\hat{O}_{\mathbf{r}_i}$ are printed blue.}
\label{DefF}
\end{figure}

{\em {\bf\em Theorem 2}: For extra terms creating 2D fractons through membrane of $\hat{O}_\mathbf{r}$ operators in the FTO, the critical point $v_{c2}$ of the phase transition satisfies $v_{c2} \leq  v_{c2}^{(u)}=3w/16$. For extra terms creating 3D fractons through body operators, the critical point $v_{c3}$ satisfies $v_{c3} \leq v_{c3}^{(u)}=19w/256$. Here $w$ is the energy of a group of 4 (or 8) fractons of the 2D membrane (3D cube).}

In particular, we show for the 2D fracton case, 
the growth of $\mathcal{H}_{\rm eff, gg'}^{(N+r)}$ is $16v/3w$ as $r=\zeta N$ increases by one order when $\left|g\right>$ and $\left|g'\right>$ is connected by a 2D membrane of operators $\hat{O}_\mathbf{r}$. This can be seen by noticing that at any point $\bold r_0$ (on which a 2D fracton is defined) that are surrounded by four operators $\hat{O}_{\mathbf{r}_1}$, $\hat{O}_{\mathbf{r}_2}$, $\hat{O}_{\mathbf{r}_3}$ and $\hat{O}_{\mathbf{r}_4}$ (listed in clockwise direction, as shown in Fig. \ref{DefF}(a)), the number of fractons $F_{2,\mathbf{r}_0}$ at $\bold r_0$ can be written as
\begin{equation}
F_{2,\mathbf{r}_0} = \sum_{i=1}^4\bigr( s_i-2s_is_{i+1}+2\prod_{j\neq i}s_j\bigr)-4\prod_i s_i,
\end{equation}
where $s_5=s_1$ and $s_i=(1+\hat{O}_{r_i})/2$. As shown through random walk theory in section~\ref{HOC}, the growth of high order contributions with infinite $r$ is proportional to the average number $\bar{F}_{2,\mathbf{r}_0}$ of excitations over all possible configuration. Note that each term of $s_i$ contributes a factor of $1/2$. It then follows that that $\bar{F}_{2,\mathbf{r}_0}=3/4$. Therefore, the growth required is $4 \times 4v/3w = 16v/3w$, giving the upper limit of the critical value $v_{c2} \leq  v_{c2}^{(u)}=3w/16$.

The upper limit of the critical point can be obtained for the case with 3D fractons in a similar way. At any point $\bold r_0$ that is surrounded by eight $\hat{O}_{\mathbf{r}_i}$ operators (as shown in Fig. \ref{DefF}(b)), the number of fractons $F_{3,\mathbf{r}_0}$ at $\bold r_0$ is given by
\begin{equation}
F_{3,\mathbf{r}_0} = \sum_{i_1 \cdots i_8} a_{i_1\cdots i_8} \prod_j s_j^{i_j}.
\end{equation}
Here $a_{i_1\cdots i_8}$ are coefficients whose values can be found in~\cite{supplementary}. The average number of fractons is then obtained that $\bar{F}_{3,\mathbf{r}_0}=29/32$, and the growth of $\mathcal{H}_{\rm eff, gg'}^{(N+\zeta N)}$ for infinite $r$ is $256v/29w$ as $r$ increases by one. Thus the upper critical point reads $v_{c3}^{(u)}=29w/256$. An external term of strength $v>v_{c2,c3}^{u}$ drives the FTO into a trivial phase.



{\em Lower limits.}--There remains an interesting issue on the lower limits of the critical point of the phase transition when the extra term can create higher dimensional (2D or 3D) fractons. The lower limit lies in identifying the growth of the leading-order contribution, whose exact solution is currently not available. However, we provide below a reasonable conjecture of the lower limits.

{\em {\bf \em Conjecture}: The critical points satisfy $v_{c2}\geq w/16$ and $v_{c3}$ satisfies $v_{c3}\geq w/64$ when the extra term can create 2D and 3D fractons, respectively.} The lower limits can be approached by estimating the maximum value of the transition matrix element between the ground states. The conjecture is based on the observation that for the 2D fractons, the number of configurations connecting two ground states in the same order perturbation is more than the case for 1D fractons due to the higher dimension in the operations. This is already seen in the exact study of the upper limits. The maximum number of the configurations could be the square of that for the case with 1D fractons, namely the square of the Calatan number as shown in the case with lineons. 
In this case, the lower limit of the critical point is given by $v_{c2}^{(l)} = w/4^2=w/16$. In the similar way, we conjecture that the lower limit for the case of 3D fractons is given by $v_{c2}^{(l)} = w/64$. The conjecture is clearly supported by numerical results, as given in the Supplementary Material \cite{supplementary}. To analytically confirm the lower limits and further the exact result of the phase transition critical point in the generic case deserves future great efforts.


\section{Conclusion}
We have presented an analytic theory of the quantum phase transition in FTO induced by external terms which create fracton excitations. We developed an effective Hamiltonian approach which is model independent to study the critical point, beyond which the ground state degeneracy (GSD) breaks down. This approach can be mapped to the random walk problem, allowing to study the critical point of quantum phase transition induced by fractons of different dimensions. For the lineonic external term, we provide a universal result of the critical point, given by $v_{c1}=w/4$ if $w_i\geq w$ or $v_{c1}' \approx \frac{w}{4}-\frac{w-w_i}{4}\times 0.26$ for $w_i<w$, with $w$ being the energy of a pair of lineons and $w_i$ being the energy of two lineons meeting at the same point. This result can be directly applied to X-cube model with transverse Zeeman field~\cite{Checkerboard}. For the external terms creating 2D or 3D fractons, e.g. the transverse field in the HHB model~\cite{SpinChain}, we show an upper limit and conjectured a reasonable lower limit with numerical verification for the critical phase transition point. Beyond the upper limit (Below the lower limit) the FTO breaks down (survives) in general. The analytic theory presented here shows insights into the understanding of the quantum phase transition in FTOs, and can be applied to predicting the full phase diagram of generic FTOs, which may facilitate the physical realization of such exotic orders.

\begin{acknowledgments}
This work was supported by National Natural Science Foundation of China (Grant No. 11825401), the Open Project of Shenzhen Institute of Quantum Science and Engineering (Grant No.SIQSE202003), and the
Strategic Priority Research Program of Chinese Academy of Science (Grant No. XDB28000000).
\end{acknowledgments}

\begin{appendix}

\section{$n^{\rm th}$ order Degenerate Perturbation Theory}

In this Appendix, we present the rigorous calculation of arbitrary $n^{\rm th}$ order degenerate perturbation on the many-body ground states, whose degeneracy is assumed to be unchanged up to the $n-1^{\rm th}$ order. Let $H = H_0 + \bar{v} V$, where $\bar{v} \ll 1$, $H_0$ denotes the unperturbed Hamiltonian, and $V$ is the external perturbation of the system. Write down that $H_0 = H_{0,gg}+H_{0,\bar{g}\bar{g}}$ and the external term $V = V_{gg} + V_{\bar{g}\bar{g}} + V_{g\bar{g}} + V_{\bar{g}g}$. The subscripts $gg$ (or $\bar{g}\bar{g}$) means that the corresponding operator is projected onto the ground state subspace (or excited state subspace). Similarly, the subscripts $g\bar{g}$ and $\bar{g}g$ mean that the operators connect the ground and excited state subspaces.

Denote by $P$ the projection operator onto the ground state subspace whose energy is $E$. For the total Hamiltonian $H$, the secular equation is
\begin{widetext}
\begin{eqnarray}
0&=& \text{det}\left(H_0 + \bar{v} V - E\right) \nonumber\\
&=& \text{det}\left[\left(H_{0,\bar{g}\bar{g}}+\bar{v} \left(V_{\bar{g}\bar{g}}+V_{g\bar{g}}\right)-E (1-P)+P\right)\times\right. \nonumber\\
&&\left.\left(1-P+\bar{v}\left(1-\bar{v} V_{g\bar{g}}\right)\left(H_{0,\bar{g}\bar{g}}+\bar{v} V_{\bar{g}\bar{g}} - E(1-P)\right)^{-1}_e V_{\bar{g}g}+H_{0,gg}+\bar{v} V_{gg} - EP\right)\right] \nonumber\\
&=& \text{det} \left(H_{0,gg}+\bar{v} V_{gg}-E-\bar{v}^2 V_{g\bar{g}} \left(H_{0,\bar{g}\bar{g}} + \bar{v} V_{\bar{g}\bar{g}} -E\right)^{-1}_e V_{\bar{g}g} \right)_g \text{det}\left(H_{0,\bar{g}\bar{g}}+\bar{v} V_{\bar{g}\bar{g}}-E\right)_e \nonumber\\
&=& \text{det} \left(H_{0,gg}+\bar{v} V_{gg}-E-\bar{v}^2 V_{g\bar{g}} \left(\frac{1}{E-H_{0,\bar{g}\bar{g}}}+\frac{1}{E-H_{0,\bar{g}\bar{g}}}\bar{v} V_{\bar{g}\bar{g}}\frac{1}{E-H_{0,\bar{g}\bar{g}}}+\cdots\right) V_{\bar{g}g} \right)_g \nonumber\\
&& \times\text{det}\left(H_{0,\bar{g}\bar{g}}+\bar{v} V_{\bar{g}\bar{g}}-E\right)_e.
\end{eqnarray}
\end{widetext}
Then the effective Hamiltonian projected on the ground state subspace is given by:
\begin{eqnarray}
H_{\rm eff} &=& H_{0,gg}+\bar{v} V_{gg}-\bar{v}^2 V_{g\bar{g}} \biggr(\frac{1}{E-H_{0,\bar{g}\bar{g}}}\nonumber\\
&&+\frac{1}{E-H_{0,\bar{g}\bar{g}}}\bar{v} V_{\bar{g}\bar{g}}\frac{1}{E-H_{0,\bar{g}\bar{g}}}+\cdots\biggr) V_{\bar{g}g}.
\end{eqnarray}
With the power series $E =\sum_j E_j \bar{v}^{j+1}$ we obtain that
\begin{eqnarray}
\frac{1}{E-H_{0,\bar{g}\bar{g}}} &=& \frac{1}{E_0-H_{0,\bar{g}\bar{g}}}-\bar{v} \frac{E_1}{\left(E_0-H_{0,\bar{g}\bar{g}}\right)^2}+\nonumber\\
&&\bar{v}^2 \left(\frac{E_1^2}{\left(E_0-H_{0,\bar{g}\bar{g}}\right)^3} - \frac{E_2}{\left(E_0-H_{0,\bar{g}\bar{g}}\right)^2}\right)+\cdots\nonumber\\
&=& \frac{1}{E_0-H_{0,\bar{g}\bar{g}}} + \sum_{i=1}^{\infty} \sum_{s\in P_i} \frac{(-1)^{\left|s\right|}\bar{v}^i \prod_{j } E_{s_j}} {\left(E_0-H_{0,\bar{g}\bar{g}}\right)^{\left|s\right|+1}},\nonumber
\end{eqnarray}
where $E_0$ is the energy for the unperturbed ground state, $P_i$ is the set of integer partitions of the integer $i$, and the partition with different orders are counted different. For example, $1+2=2+1=3$ are two different partitions of $i=3$. In the summation $s_j$ is the $j^{\rm th}$ term in the integer partitions $s$ and $\left|s\right|$ is the number of integers partitioned. For example, if $s=3+1+5+4\in P_{13}$, one has that $s_3 = 5$, $s_4=4$, and $\left|s\right|=4$). For convenience, we further rewrite that $(E-H_{0,\bar{g}\bar{g}})^{-1}=\sum_{i=0}^\infty \bar{v}^i Q^{(i+1)}$, with $Q^{(i)}$ in the expansion given by
\begin{equation}\label{Eqn:Qi}
Q^{(i)} = - \sum_{s\in P_{i-1}} \frac{\prod_{j } E_{s_j}} {\left(H_{0,\bar{g}\bar{g}}-E_0\right)^{\left|s\right|+1}}.
\end{equation}
For example, the $4^{\rm th}$ order term $Q^{(5)}$ of $\left(E-H_{0,\bar{g}\bar{g}}\right)^{-1}$ is $E_1^4/Q_0^5 - 3E_1^2E_2/Q_0^4 + E_2^2/Q_0^3 + 2E_3E_1/Q_0^3 - E_4/Q_0^2$, where we have defined $Q_0\equiv\left(Q^{(1)}\right)^{-1} = E_0-H_{0,\bar{g}\bar{g}}$. With these results the effective Hamiltonian projected onto the ground state subspace can be further simplified in the following way
\begin{widetext}
\begin{eqnarray}
H_\text{eff} &=& H_{0,gg} + \bar{v} V_{gg} - \bar{v}^2 V_{g\bar{g}} \frac{1}{Q_0}V_{\bar{g}g} - \bar{v}^3 V_{g\bar{g}} \left(\frac{1}{Q_0}V_{\bar{g}\bar{g}}\frac{1}{Q_0}+Q^{(2)}\right)V_{\bar{g}g} \nonumber\\
&& - \bar{v}^4 V_{g\bar{g}} \left(\frac{1}{Q_0}V_{\bar{g}\bar{g}}\frac{1}{Q_0}V_{\bar{g}\bar{g}}\frac{1}{Q_0}+Q^{(2)}V_{\bar{g}\bar{g}}\frac{1}{Q_0}+\frac{1}{Q_0}V_{\bar{g}\bar{g}}Q^{(2)}+Q^{(3)}\right)V_{\bar{g}g} - \cdots \nonumber\\
&=& H_{0,gg} + \bar{v} V_{gg} - \sum_{i=2}^{\infty} \bar{v}^i V_{g\bar{g}} \left( \sum_{(s_1,\cdots, s_m)=s \in P_{i-1}} Q^{(s_1)} V_{\bar{g}\bar{g}} Q^{(s_2)} V_{\bar{g}\bar{g}} \cdots V_{\bar{g}\bar{g}} Q^{(s_m)} \right) V_{\bar{g}g}.  \label{Eqn:DegPerEffH}
\end{eqnarray}
\end{widetext}

\end{appendix}

\onecolumngrid
\newpage
\centerline{\bf \large Supplementary Material: Quantum phase transition of fracton topological orders}

\beginsupplement
\tableofcontents
\newcounter{TOCref}
\setcounter{TOCref}{19}
\newcommand{\editcontentsline}[3]{\addtocontents{#1}{\protect\contentsline{#2}{#3}{\hspace*{-5pt}\thepage}{section*.\theTOCref}}\refstepcounter{TOCref}}
\newcommand{\tocless}[2]{\bgroup\let\addcontentsline=\editcontentsline#1{#2}\egroup}
\let\oldsection\section
\let\oldsubsection\subsection
\let\oldsubsubsection\subsubsection
\renewcommand{\section}[1]{\tocless\oldsection{#1}}
\renewcommand{\subsection}[1]{\tocless\oldsubsection{#1}}
\renewcommand{\subsubsection}[1]{\tocless\oldsubsubsection{#1}}

\section{General properties of the perturbation theory}
Not all terms in the perturbation theory are meaningful. The following section provides a method to represent the terms diagrammatically, so that one can clearly see which terms in the perturbation series are valid and which are not.

\subsection{Diagrammatic representation of perturbation terms}
\begin{figure}[!ht]
\centerline{\includegraphics[width=.9\columnwidth]{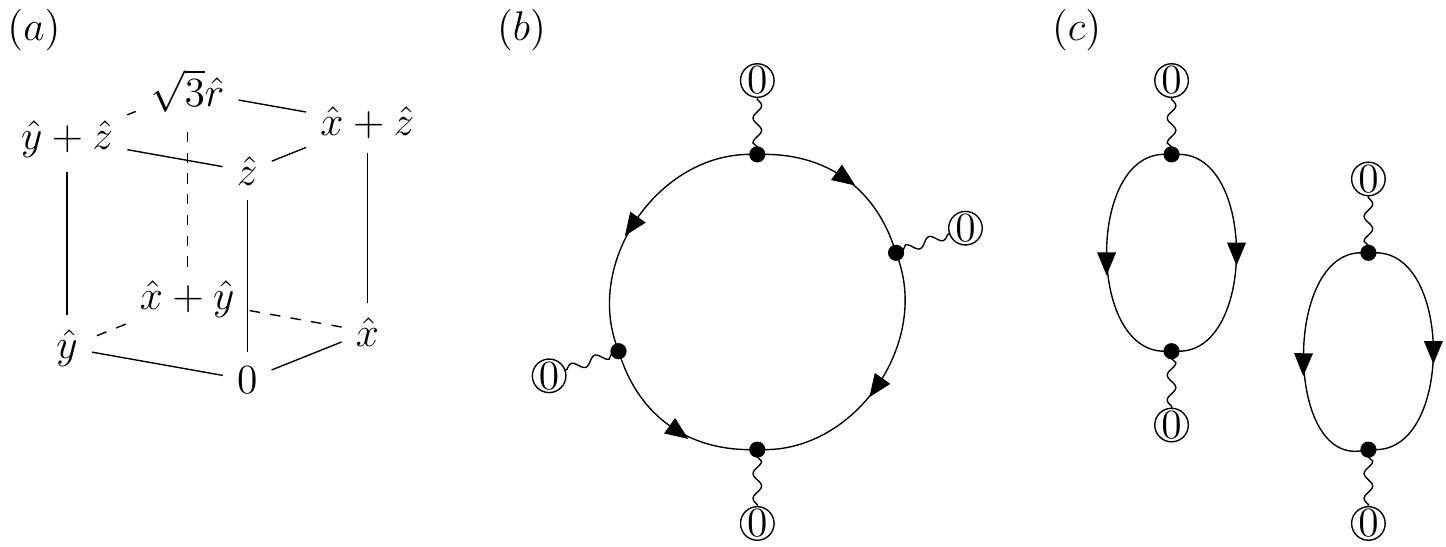}}
\caption{An example of employing the diagrammatic rule on an FTO system. (a) The basic configuration of the lattices in FTO system. In particular, if periodic boundary condition is taken with size $N$, $N\hat{x}=N\hat{y}=N\hat{z}=0$ (b) A diagram representing an off-diagonal term. (c) A diagram representing an diagonal term.}
\label{Fig:ProofH1}
\end{figure}
We will first present the diagrammatic rules and gives some examples to clarify the rules. (i) The diagrams are read from top to bottom. (ii) The curvy lines represent the interaction $V$ and thus the total number of the curvy lines in a graph is the order of the perturbation. (iii) The vertexes connecting the solid lines and curvy lines are called internal vertexes. They represent three possibilities when $V$ is applied on the state. When two solid lines are connected from above, two lineons are destroyed; when one is connected from above and one from below, one lineon is moved; when two are connected from below, two lineons are created. (iv) The vertexes connecting only the curvy lines are called external vertexes. (v) The solid lines represent the flow of lineons. (vi) Starting from the top internal vertex to the bottom internal vertex, no horizontal line can be drawn without cutting any solid lines. (vii) Loops with one internal vertex is prohibited. (viii) Within the diagram, the parts with same external vertexes are called a subdiagram. A horizontal line that cuts through any subdiagrams is called transitional line and the lines are labeled with positive numbers distinct from other transitional lines. (The number $0$ is reserved for main diagram.) (ix) That number are used to construct the external vertexes of a new subdiagram.

Rules 1 to 5 are conventions to denote a specific perturbation term. The $6^{\rm th}$ rule is to guarantee that the transition must be from ground state to ground state, and that all intermediate states are excited states. If a horizontal line can be drawn, at that point the state transforms back to the ground state, which is prohibited (c.f. Appendix). The $7^{\rm th}$ rule is to forbid a simultaneous creation and annihilation of a lineon. With this seven rules, it is possible to consider the terms in the form $\bar{v}^n V_{g\bar{g}} \left(Q_0^{-1} V_{\bar{g}\bar{g}}\right)^{n-2} Q_0^{-1} V_{\bar{g}g}$. In particular, Fig. \ref{Fig:ProofH1} shows an example. (a) shows a unit cell of a simple cubic lattice with size $N^3$ with periodic boundary condition. Suppose $N=4$, then (b) may express terms like $\bar{v}^4 \hat{O}_{\hat{y}}  Q_0^{-1} \hat{O}_{0} Q_0^{-1} \hat{O}_{2\hat{y}} Q_0^{-1} \hat{O}_{3\hat{y}}$. Note that the leading order perturbations must be drawn with a single loop and describe a non-contractible loop in FTO. For diagrams with more than one loop, for example Figure (c), it may express terms like $\bar{v}^4 \hat{O}_{2\hat{y}}  Q_0^{-1} \hat{O}_{\hat{x}} Q_0^{-1} \hat{O}_{2\hat{y}} Q_0^{-1} \hat{O}_{\hat{x}}$, so that the lineons at the neighborhood of $2\hat{y}$ and $\hat{x}$ just annihilate with their partners nearby without mutual communication. In addition, it is worth noting that in the diagrammatic language if the smallest non-contractible loop has $N$ sites, there must be at least a loop of order at least $N$ (i.e. with at least $N$ vertexes) to connect different ground states. Such loops are called {\it main loops}. Therefore, among two diagrams, diagram (b) represents off-diagonal terms in the effective Hamiltonian.

\begin{figure}[!ht]
\centerline{\includegraphics[width=.5\columnwidth]{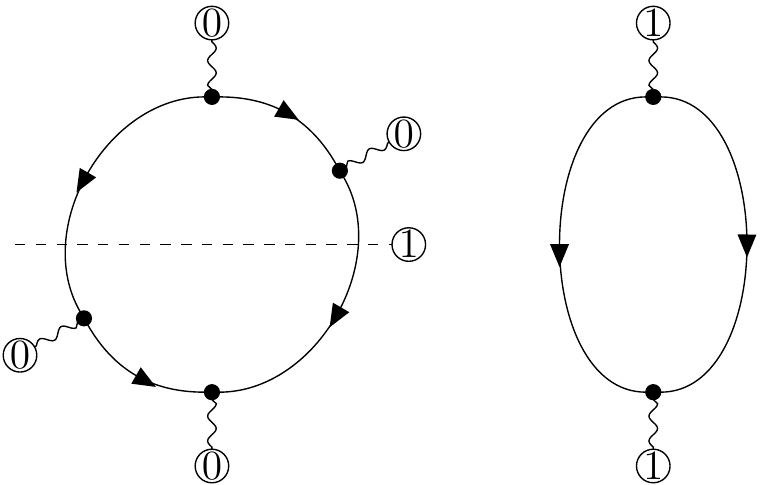}}
\caption{An example for diagram with a transitional line.}
\label{Fig:ProofH2}
\end{figure}

More generally, for the terms of the form $\bar{v}^n V_{g\bar{g}} \left(\Pi_{i=1}^{r-1} Q^{(\alpha_i)} V_{\bar{g}\bar{g}}\right) Q^{(\alpha_r)} V_{\bar{g}g}$, where some of the $\alpha_i$ are larger than $1$ and $(\alpha_1 \cdots \alpha_r) \in P_{n-1}$, the last two rules (viii) and (ix) are needed. The transitional lines and subdiagrams are to record the places at which $Q$ appear and the corresponding $E_s$ terms as in the Appendix. For example, a sixth order term is shown pictorially in Fig. \ref{Fig:ProofH2}. The diagram represents a term $\bar{v}^6 V_{g\bar{g}} Q^{(1)} V_{\bar{g}\bar{g}} Q^{(3)} V_{\bar{g}\bar{g}} Q^{(1)} V_{\bar{g}g}$, where $Q^{(3)'} = -E_2/Q_0^2$, where the minus sign is determined by the last rule.

To calculate contribution from the diagram, three calculation rules are needed. (i) Within each subdiagram, just below each internal vertexes (except the bottom one) or transitional lines, imagine a horizontal line that cut through the diagram and let $\zeta$ to be the number of solid lines it cuts. Give a factor $-(\zeta w/2)^{-1}$ for each counting. (This is just to record the contribution by $\left(E_{0B}-H_{0,\bar{g}\bar{g}}\right)^{-1}$) (ii) Multiply the factor $v^C$, where $C$ is the number of curvy lines. (iii) Multiply the factor $(-1)^F$, where $F$ is the total number of transitional lines.
These calculation rules are evident for the effective Hamiltonian recorded in Appendix.

\subsection{Simplification of perturbation series}
Consider a Feymann's diagram $D$ in which there is a main loop and other loops (called auxiliary loops). Physically, the main loop describes the process to connect two different ground states, and the auxiliary loop should not have any effect on the physical result since they are spacially separated. This physical requirement are called the cluster decomposition principle. This physical properties can be transformed into diagrammatic language as follows: given a diagram $D$ with $n>1$ loops, draw a family of diagrams $\widetilde{D}$ in the following step. (i) Add arbitrary transitional lines and assign that number on the external vertexes of at least one loop. (ii) If any subdiagram contains more than one loop, stretch the loops arbitrarily so that the internal vertices can be of different order by having different relative height. (iii) All diagrammatic rules are still enforced. For example, the $6^{\rm th}$ rule -- when stretching the loops of the subdiagrams with more than one loop, there should not be a horizontal line that can pass through a subdiagram without intersecting any solid lines. Then, we claim that

\begin{figure}[t]
\centerline{\includegraphics[width=.8\columnwidth]{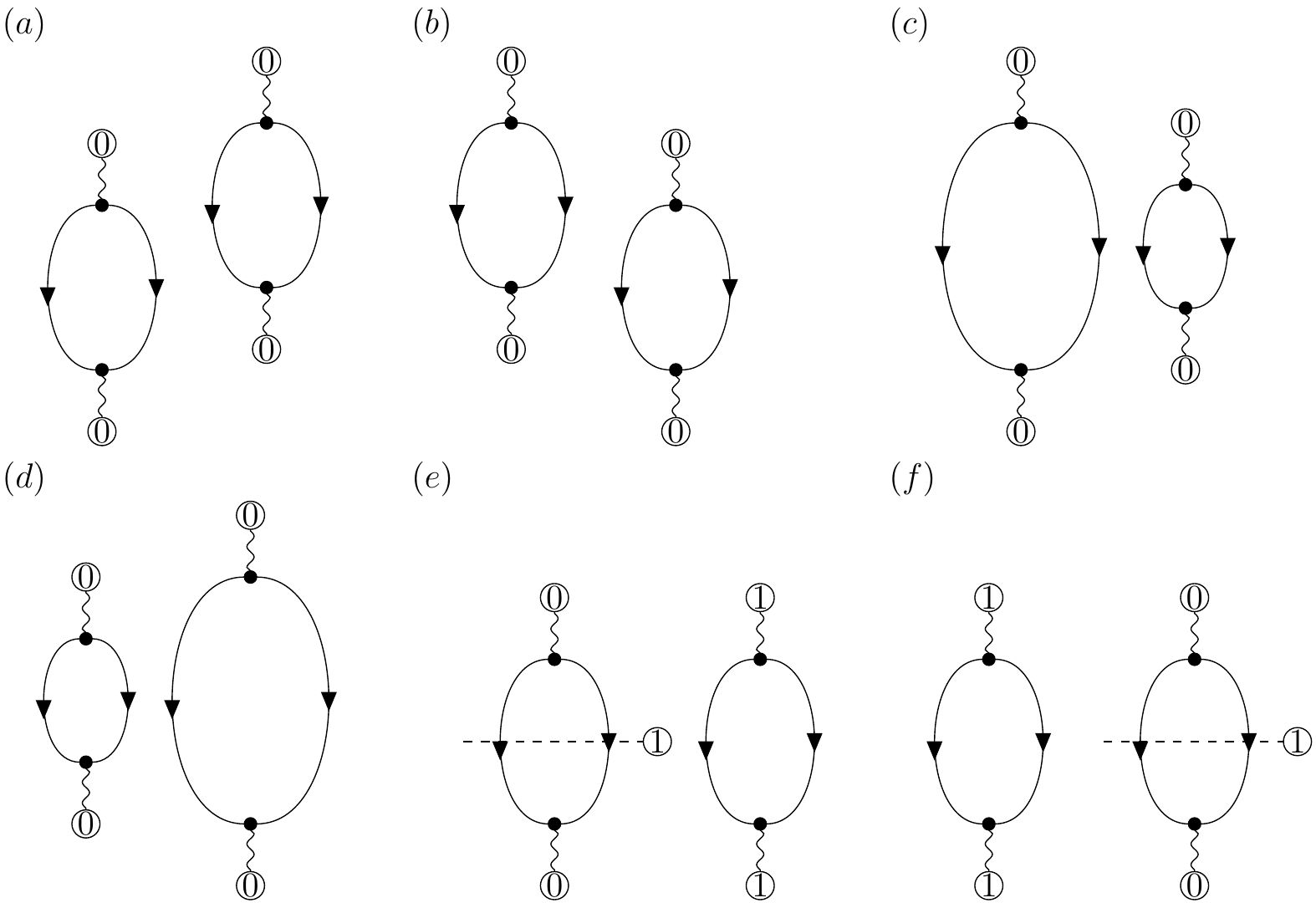}}
\caption{All diagrams in the family of diagrams $\widetilde{D}$, such that $D$ is a diagram with two loops of order 2.}
\label{Fig:ProofH3}
\end{figure}

{\em {\em\bf Claim}: Given a diagram with $n>1$ loops, the sum of the contribution of a family of diagrams containing such diagram is zero. Label the proposition for the case with $n$ loops and each loop have $r_i+1$ internal vertexes to be $P_n(r_1,\cdots,r_n)$.} Take the proposition of $P_2(1,1)$ as an example (Fig. \ref{Fig:ProofH3}). Assume at the moment that the left loop produce excitations with energy of $E_a$ and the right loop $E_b$. Label the sum of the family of diagrams as $S_2(1,1)$. Then
\begin{eqnarray}
S_2(1,1) &=& \frac{1}{E_b (E_a+E_b) E_a} + \frac{1}{E_a (E_a+E_b) E_b} + \frac{1}{E_a (E_a+E_b) E_a} \nonumber\\
&=& 0.
\end{eqnarray}

Since a diagram with more than $2$ loops can be decomposed into $2$ groups, it is actually suffix to prove $P_2(n,m)$. So the proof can be done by mathematical induction. Note here that $P_2(1,1)$ is explicitly shown. Suppose now $P_2(n\geq 1,1)$ is correct. Denote $a_i$ to be the energy under the $i^{\rm th}$ internal vertex for the order $n$ loop and $b$ is the energy of the order 2 loop. The sum of the family of diagrams $S_2(n,1)$ is
\begin{eqnarray}
S_2(n,1) &=& \frac{1}{b^2} \frac{1}{(a_1+b)\cdots(a_n+b)}+\sum_{i=1}^n \frac{1}{b} \frac{1}{(a_1+b)\cdots(a_i+b)a_i\cdots a_n} \nonumber\\
&& + \sum_{i=1}^n \frac{1}{b} \frac{1}{a_1\cdots a_i(a_i+b)\cdots (a_n+b)} \nonumber\\
&&+\sum_{i_2=i_1}^n \sum_{i_1}^n \frac{1}{a_1\cdots a_{i_1} (a_{i_1}+b)\cdots(a_{i_2}+b)a_{i_2}\cdots a_n} - \frac{1}{a_1\cdots a_n} \left(\sum_i^n \frac{1}{a_i}\right)\frac{1}{b}\nonumber\\
&=& 0.
\end{eqnarray}
Denote the $i^\text{th}$ term of $S_2(n,1)$ be $q_{i,n}$, then
\begin{eqnarray}
S_2(n+1,1) &=& q_{1,n} \frac{1}{a_{n+1}+b} + q_{2,n} \frac{1}{a_{n+1}} + \frac{1}{b} \frac{1}{(a_1+b)\cdots(a_{n+1}+b)a_{n+1}} \nonumber\\
&&+q_{3,n} \frac{1}{a_{n+1}+b} + \frac{1}{b} \frac{1}{a_1\cdots a_{n+1}(a_{n+1}+b)} + q_{4,n} \frac{1}{a_{n+1}} \nonumber\\
&&  + \sum_{i=1}^{n+1} \frac{1}{a_1\cdots a_i (a_i+b)\cdots (a_{n+1}+b) a_{n+1}} - q_{5,n} \frac{1}{a_{n+1}} - \frac{1}{a_1\cdots a_n a_{n+1}^2} \frac{1}{b} \nonumber\\
&=& \left(q_{1,n}+q_{3,n}\right) \frac{-b}{a_{n+1}(a_{n+1}+b)} + q_{1,n} \frac{b}{a_{n+1}(a_{n+1}+b)} \nonumber\\
&&+\frac{1}{b}\frac{1}{a_1\cdots a_{n+1}} \frac{-b}{a_{n+1}(a_{n+1}+b)} + q_{3,n} \frac{b}{(a_{n+1}+b)a_{n+1}} \nonumber\\
&& + \frac{1}{a_1\cdots a_{n+1} (a_{n+1}+b)a_{n+1}}\\
&=& 0. \nonumber
\end{eqnarray}
Therefore, by mathematical induction, the proposition $P_2(n,1)$ is true for all $n$. A similar calculation gives induction on $m$ as well so that $P_2(n,m)$ is true for all $n$ and $m$.

The above sections tell how an arbitrary order of valid perturbation looks like. In particular, the only leading order term that survives look like
\begin{equation}
\mathcal{H}_{{\rm eff},{gg'}}^{(N)}=\bar{v}^N V_{g\bar{g}} Q^{(1)} V_{\bar{g}\bar{g}} \cdots V_{\bar{g}\bar{g}} Q^{(1)} V_{\bar{g}g},
\end{equation}
where there are $N$'s $V$ in the equation, since all other terms that are of the same order have fewer $V$ so that they together cannot form a non-contractible loop. On the other hand, the appearance of $Q^{(s>2)}$ in the perturbation series makes the diagram contain subdiagrams so that it is canceled by all possible combinations of such kind of diagram (as shown in the last section). Therefore, the higher order term can also be written in a similar form
\begin{equation}
\mathcal{H}_{{\rm eff},{gg'}}^{(N+r)}=\bar{v}^{N+r} V_{g\bar{g}} Q^{(1)} V_{\bar{g}\bar{g}} \cdots V_{\bar{g}\bar{g}} Q^{(1)} V_{\bar{g}g},
\end{equation}
where there are $N+r$'s $V$ in the equation.\\

\section{Critical point of phase transition with 1D fractons}\label{MapRW}

\subsection{Calculation of the leading order perturbation}

To break the ground state degeneracy, the effective Hamiltonian must have non-zero off-diagonal elements. In particular, given the external term as $V = v \sum_\mathbf{r} \hat{O}_\mathbf{r}$, the matrix element of leading order is $\left<g\left| H_{\rm eff}^{(N)} \right|g'\right>$, where $\left|g\right>$ and $\left|g'\right>$ differs only by a global line operator. Denote $\alpha_i$ to be the place where $\hat{O}_i$ is defined so that $\prod_i \hat{O}_{i}$ connects two ground states. Write $V_{\gamma\gamma'} = \left<\gamma\left|V\right|\gamma'\right>$, then
\begin{eqnarray}
\left<g\left|H_{\rm eff}^{(N)}\right|g'\right> &=& \sum_{\beta_1\beta_2\cdots \beta_{N-1}}\left(\frac{-1}{w}\right)^{N-1} \frac{V_{g\beta_1}V_{\beta_1\beta_2}\cdots V_{\beta_{N-1} g'}}{F_{\beta_1}F_{\beta_2}\cdots F_{\beta_{N-1}}} \nonumber\\
&=& v \left(\frac{-v}{w}\right)^{N-1} {\sum_{\beta_1\beta_2\cdots \beta_{N-1}}}' \frac{1}{F_{\beta_1}F_{\beta_2}\cdots F_{\beta_{N-1}}}, \label{GSmatrixelement}
\end{eqnarray}
where $V_{\beta_i \beta_j}$ is the matrix element $\left<\beta_i\left| V \right| \beta_j \right>$ and is equal to either equal to $v$ or 0, $F_{\beta}$ is the number of pairs of lineons at the state $\beta$ and the prime in the summation sign means the summation over intermediate states are limited to those with $V_{\gamma\gamma'}=v$. Let $v_{ps}$ be the result of the primed summation in (\ref{GSmatrixelement}) . Then

{\em {\bf\em Claim}: If $N$ is the system size, then
\begin{equation}
v_{ps}(N) = {\sum_{\beta_1\beta_2\cdots \beta_{N-1}}}' \frac{1}{F_{\beta_1}F_{\beta_2}\cdots F_{\beta_{N-1}}} = C^{N-1}_{2N-2}.\label{Eqn:ClaimLeadingOrder}
\end{equation}}
If the claim is true, then as $N\rightarrow \infty$, the asymptotic behavior of the matrix element is
\begin{equation}
\left<g\left|H_{\rm eff}^{(N)}\right|g'\right> = v\left(\frac{-v}{w}\right)^{N-1} C^{N-1}_{2N-2} \sim  v\left(\frac{-v}{w}\right)^{N-1} \frac{4^{N-1}}{\sqrt{\pi N}} \sim \left(\frac{4v}{w}\right)^{N-1}.
\end{equation}
This value is zero whenever $4v/w <  1$ and thus $v_c = w/4$ to this order.

\subsubsection{Exact calculation of the leading order perturbation}
\begin{figure}[ht]
\centerline{\includegraphics[width=.9\columnwidth]{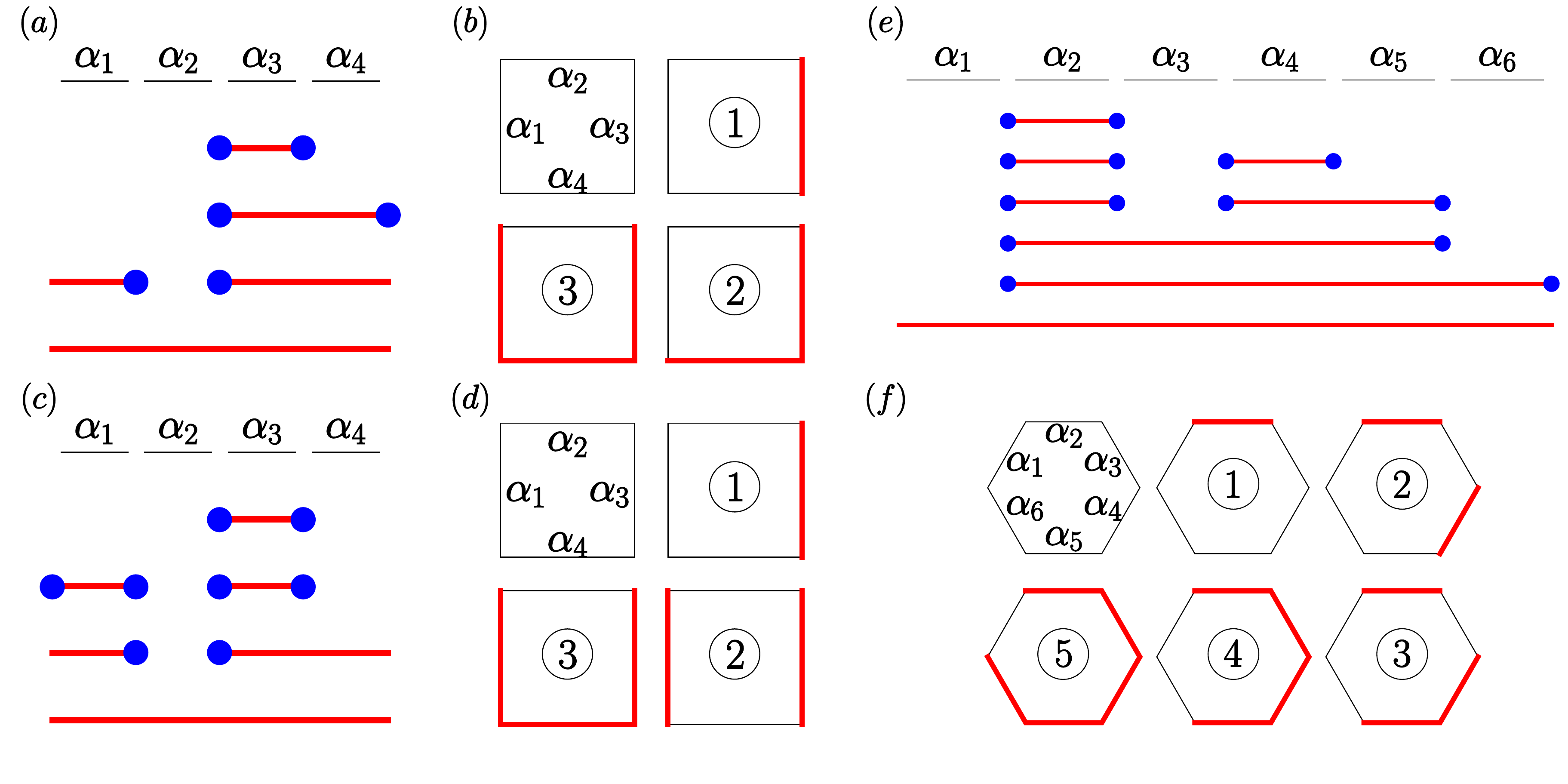}}
\caption{Three examples (a,c,e) for the meaning of $F_{\beta_i}$ and the order of application of $\hat{O}$, where the order is read from top to bottom. Note that (a) and (b), (c) and (d) and (e) and (f) correspond to each other. (a,b) $N=4$ and the corresponding order of $\hat{O}$ is $\hat{O}_3$, $\hat{O}_4$, $\hat{O}_1$, $\hat{O}_2$. The number of lineons that are produced after every action is 2. Similarly, one can interpret it as painting the edges $\alpha_3$, $\alpha_4$ and $\alpha_1$ one at a time. After each action, there are only 1 line segment. In this case, $\prod_i F_{\beta_i} = 1 \times 1 \times 1 = 1$. (c,d) $N=4$ and the order is $\hat{O}_3$, $\hat{O}_1$, $\hat{O}_4$, $\hat{O}_2$. The number of lineons that are produced successively are $2,4$ and $2$, respectively. Or, equivalently, there are $1,2$ and $1$ line segments, respectively. In this case, $\prod_i F_{\beta_i} = 1 \times 2 \times 1 = 2$. (e,f) $N=6$ and the order is $\hat{O}_2$, $\hat{O}_4$, $\hat{O}_5$, $\hat{O}_3$, $\hat{O}_6$, $\hat{O}_1$. Simiarly, the number of lineons that are produced successively are $2,4,4,2$ and $2$, respectively, or $1,2,2,1$ and $1$ line segments, respectively. So $\prod_i F_{\beta_i} = 1 \times 2 \times 2 \times 1 \times 1 = 4$.}
\label{Fig:Proof1}
\end{figure}

Suppose there are $N$ lattices $\alpha_1, \alpha_2, \cdots, \alpha_N$ so that the transition occurs when $\prod_{i=1}^N \hat{O}_i$ is applied on $\left|g'\right>$. For leading order perturbation, the sum is actually over permutations group of $\alpha_1, \alpha_2, \cdots, \alpha_N$. The permutation corresponds to the order of how $\hat{O}_i$ are applied on the ground state $\left|g'\right>$, as mentioned in the main text (or described in Fig. \ref{Fig:Proof1}), so
\begin{equation}
v_{ps}(N) = \sum_{p\in S_N} \frac{1}{\widetilde{F}_{p_1} \widetilde{F}_{p_2} \cdots \widetilde{F}_{p_{N-1}}},\label{Eqn: summand}
\end{equation}
where $p=(p_1,p_2,\cdots, p_{N-1}, p_N) \in S_N$ and $\widetilde{F}_{p_i}$'s are the number of pairs of lineons after applying $\hat{O}$ on the lattices $\alpha_{p_1}, \alpha_{p_2}, \cdots, \alpha_{p_i}$. Thus for every permutation $p$, we can define the following angle notation $p_{<>} = \left< a_1, a_2, \cdots, a_i, \cdots , a_{N-1} \right>$. The notation with $N-1$ numbers refers to a permutation $p \in S_N$. The numbers $a_i$'s mean that there are $a_i$ pairs of lineons after the operation $\prod_{j=1}^i \hat{O}_{p_j}$. For example, for the examples shown in the Fig. \ref{Fig:Proof1}(a,c,e), the angle notations of the permutations are $\left<1,1,1\right>$, $\left<1,2,1\right>$ and $\left<1,2,2,1,1\right>$, respectively. All permutations with the same angle notations are called equivalent permutation. Below are the statements and proofs of two claims, which count the number of equivalent permutations in two different cases.

{\em Counting Equivalent Permutations (I)}---.Note that $\left|a_{i+1} - a_{i} \right|\leq 1$, since an application of $\hat{O}$ on a lattice either creates ($1$) or destroys ($-1$) two lineons or moves one lineon to another place ($0$). Here, we shall first consider the case that the angle notation satisfies $a_{i+1} - a_i \neq 0$. In this case, there are $N \prod a_i$ equivalent permutation. To prove this statement, suppose the first element of any permutation is $1$ so that at the end of calculation, the equivalent permutation should be multiplied by $N$. Now, there are three different cases:

(i) Suppose $p_{<>}$ is of the form $\left<1,2,\cdots, n-1, n, n-1, \cdots, 1 \right>$. Then the first $n$ terms in $p \in S_{N = 2n}$ must be odd so that there are in total $(n-1)!$ ways to paint the first $n$ edges and there are also $n!$ ways to paint the last $n$ edges. Therefore the total number of equivalent permutation is $2n \times n! (n-1)! = N \prod_i a_i$.

(ii) Suppose $p_{<>}$ is of the form $\left<1,\cdots, n, n-1 \cdots, m, m+1,\cdots, p ,p-1 ,\cdots, 1\right>$ (i.e. $a_i$'s increase from $1$ to $n$, decrease from $n$ to $m$, increase from $m$ to $p$, and finally decrease from $p$ to 1). Here, whenever $a_i-a_{i-1}=\pm 1$, the $p_i$ must be odd (even), otherwise there will be some $j$ that satisfy $a_j-a_{j-1}=0$. The counting will be done in four steps: (I) Similar to the first $n$ terms in the last case, and it contributes $(n-1)!$. (II) The next $n-m$ action are to decrease the number of segments. The only possible way is when they link up the segments formed in (I), and there are in total $P^{n-m}_n$ ways to do it. After the counting, we get $m$ ordered segments. (III) The next $p-m$ action are to increase the number of segments. This is the same as inserting $p-m$ odd segments in the gaps of $m$ ordered segments. The number of ways to do it is $H^{p-m}_m (p-m)!$, where $H^r_n = C^r_{n+r-1}$ is the number of ways to place $r$ identical items into $n$ categories. The factorial factor $(p-m)!$ counts also the order of application of that $p-m$ odd segments. After inserting the odd segments, all numbers $p_{1\leq i \leq 2n-2m+p}$ are identified automatically since the lengths of the segments and the relative distances of the segments to the first number are all known. (IV) the last $p$ actions and the generalization to $p_1\neq 1$ are similar to the last case and they contribute $p! \times N$. Then the number of equivalent family can be calculated as
\begin{equation}
\left| p_{<>}\right| = (n-1)! \times \frac{n!}{m!} \frac{(p-1)!}{(p-m)!(m-1)!}(p-m)! \times p! \times N = N\frac{n! (n-1)! p! (p-1)!}{m! (m-1)!} = N\prod_i a_i.
\end{equation}

(iii) Suppose $p_{<>}$ is in the most general case such that $a_i-a_{i-1} = \pm 1$. Suppose the numbers $a_i$ increase and decrease so that they peak at $n_1, \cdots, n_{r+1}$ and trough at $m_1, \cdots, m_r$ successively. Here $N = 2\sum_r\left(n_r - m_r\right)$. Following the same spirit of the last case, we split the counting into several steps, where some steps will be iterated. (I, II) are similar and they contribute $(n_1-1)! \times P^{n_1-m_1}_{n_1}$. Then the following two steps are iterated. (a) The next $n_2-m_1$ action are to increase the number of segments by adding odd segments in the list. It is similar to the first step when we treat the old $m_1$ segments as $m_1$ odd numbers with a particular ordering. Then total number of such action is $(n_2-1)!/(m_1-1)!$ (The division is present since the order  is chosen in the first step). (Make the replacements $n_2 \rightarrow n_{j+1}$, $m_1 \rightarrow m_j$ and $m_2\rightarrow m_{j+1}$ for $j^{\rm th}$ iteration process.) (b) Similar to second step, there are $P^{n_2-m_2}_{n_2}$ to connect the $n_2$ odd segment to $m_2$ segments. (Make the replacements $n_2 \rightarrow n_{j+1}$ and $m_2\rightarrow m_{j+1}$ for $j^{\rm th}$ iteration process.) (III, IV) are again similar and they contribute the factors $H^{n_{r+1}-m_r}_{m_r}\times n_{r+1}! \times N$. So
\begin{eqnarray}
\left| p_{<>}\right| &=& (n_1-1)! \times \frac{n_1!}{m_1!} \times \left(\prod_{j=1}^{r-1} \frac{(n_{j+1}-1)!}{(m_j-1)!} \times \frac{n_j!}{m_j!} \right) \times \frac{(n_{r+1-1)}!}{\left(m_r-1\right)!} \times n_{r+1}! \times N\nonumber\\
&=& N \frac{\prod_{i=1}^{r+1} (n_i-1)! n_i!}{\prod_{i=1}^{r} (m_i-1)! m_i!} = N\prod_i a_i.
\end{eqnarray}
This completes the counting of equivalent permutations so that whenever $a_{i+1}-a_i\neq 0$ for all $i$ the equivalent permutations is $N \prod_i a_i$.

{\em Counting Equivalent Permutations (II)}---.A generalization to the above case is that $a_{i+1}-a_i = 0$ for some $i$. For this, we let $q$ to be the number of such $i$. Similar to the step (a) of case (iii) in the last proof, we imagine there are $a_i$ segments, and they behave likes $a_i$ numbers with specific ordering. When $a_{i+1}-a_i = 0$, the $i+1^{\rm th}$ action is to lengthen an existing segment. Since each action of lengthening gives a factor of $2a_i$. (Each segment has two ways to lengthen - on the left and on the right.) Therefore, there are extra factors of $2a_i/a_i = 2$ for every $i$ such that $a_{i+1}-a_i = 0$. So,
\begin{equation}
\left| p_{<>}\right| = 2^q \times N \prod_i^{N} a_i.
\end{equation}
Knowing the number of equivalent permutations in the most general case, we can count $v_{ps}$ by summing over all families of equivalent permutations. Then we arrive the following result.

{\em Summing over Equivalent Permutations}---.For every permutation $p \in S_N$ with angle notation $p_{<>}$, define $f(p) = \prod_{i=1}^{N-1} a_i^{-1}$, then the formula
\begin{equation}
\sum_{p \in S_N} f(p) = C^{N-1}_{2N-2}, \label{Eqn:ClaimLeadingOrder2}
\end{equation}
is equivalent to (\ref{Eqn:ClaimLeadingOrder}). Now we try to prove this formula in order to complete the calculation of $v_{ps}$. Let $S_N'$ to be the set of families of equivalent permutations in $S_N$. Its element $\widetilde{p} \in S_N'$ means $p \in \widetilde{p}$, so $\widetilde{p}_1=\widetilde{p}_2$ implies $p_1$ and $p_2$ are equivalent. Then the summand simplifies to
\begin{equation}
\sum_{p \in S_N} f(p) = \sum_{\widetilde{p} \in S_N'} \frac{\left| p_{<>} \right| }{\prod_{i=1}^{N-1} a_i} =  N\sum_{\widetilde{p} \in S_N'} 2^{q(\widetilde{p})}, \label{Eqn:ClaimLeadingOrderSim}
\end{equation}
where $a_i$'s are the numbers in the angle notation $p_{<>}$ and $q(\widetilde{p})$ is just the number of $i$ such that $a_{i+1}-a_i = 0$ for any $p$ in the family of equivalent permutation $\widetilde{p}$. Thus $q(\widetilde{p})$ can also be represented by any element in $\widetilde{p}$, $q(p)$. Define $q_r(p)$ to be the number of $i \leq r$ such that $a_{i+1}-a_i = 0$. Then for all $\lambda$, where $1\leq \lambda \leq N-1$, the following equation holds:
\begin{eqnarray}
\sum_{\widetilde{p} \in S_N'} 2^{q(p)} = \sum_{a_1 \cdots a_{N-1}} 2^{q(p)}  &=& \left(\sum_{a_1 \cdots a_{\lambda-1}} 2^{q_\lambda(p)}\right)\left(\sum_{a_\lambda \cdots a_{N-1}} 2^{q(p)-q_\lambda(p)}\right) \nonumber\\
&=& \sum_\rho b_{\rho, \lambda} \left(\sum_{a_{\lambda+1}\cdots a_{N-1}} 2^{q(p)-q_\lambda(p)}\right)  \label{Eqn:SplittedSumLeadingOrder},
\end{eqnarray}
where the summation sign with $a_i$ means to sum over all possible values of $i^{\rm th}$ term in $p_{<>}$ and $b_{\rho, \lambda} = \sum_{a_1 \cdots a_{\lambda-1}} \left. 2^{q_\lambda(p)}\right|_{a_\lambda = \rho}$ such that the $\lambda^{\rm th}$ term of the angle notation $p_{<>}$ is $\rho$. The first equal sign is true because the summation over family of equivalent permutations is equivalent to the summation over all possible angle notations. The second equal sign is true because the exponent $2^{q(p)}$ can be factored into two parts, with each part depends only the corresponding $a_i$'s. Now, the value of the summation is just $b_{1,N-1}$ since the last element of the angle notation must be $1$. Then from the definition of $b_{\rho,\lambda}$, the following recursion relation holds
\begin{eqnarray}
b_{\rho,\lambda} = \left.\sum_{a_1 \cdots a_{\lambda-1}} 2^{q_\lambda(p)}\right|_{a_\lambda = \rho}  &=& \left.\sum_{\rho'}\left(\left.\sum_{a_1 \cdots a_{\lambda-2}} 2^{q_{\lambda-1}(p)}\right|_{a_{\lambda-1}=\rho'}\right) 2^{q_{\lambda}(p)-q_{\lambda-1}(p)}\right|_{a_\lambda=\rho} \nonumber\\
&=&  \left.\sum_{\rho'} b_{\rho',\lambda-1} 2^{q_{\lambda}(p)-q_{\lambda-1}(p)}\right|_{a_\lambda=\rho} \nonumber\\
&=& b_{\rho-1,\lambda-1} + 2b_{\rho,\lambda-1} + b_{\rho+1,\lambda-1} \label{Eqn:RecursionRelationb},
\end{eqnarray}
where the third equal sign is true because $\rho' = \rho \pm 1$ or $\rho' =\rho$, and for the last case, the summand contributes an extra factor of 2. The boundary condition are given by $b_{0,\lambda} = b_{\lambda>\rho, \rho} =0$ and $b_{1,1} = 1$.

\begin{figure}[h!]
\centerline{\includegraphics[width=.8\columnwidth]{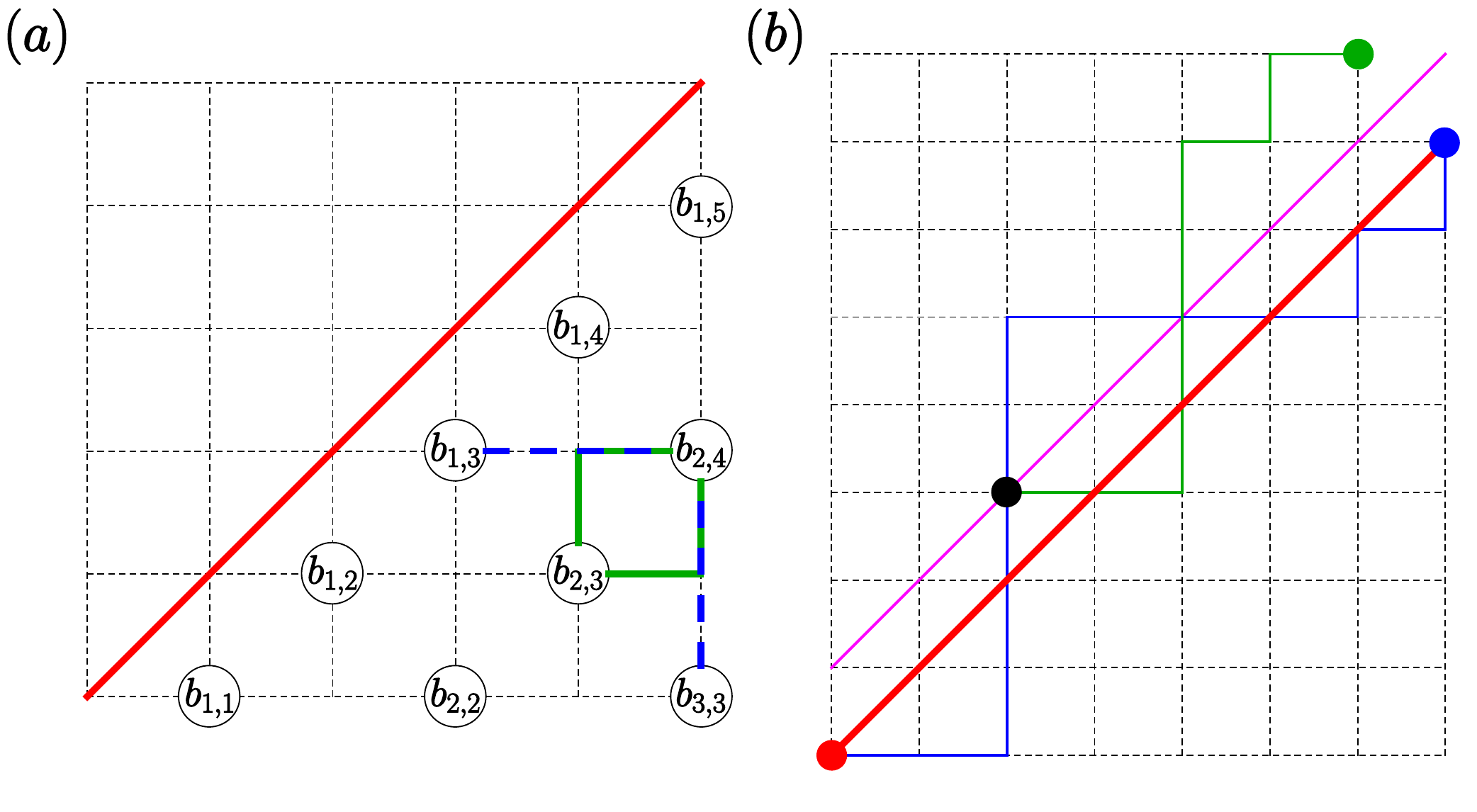}}
\caption{(a) Schematic diagram to show the matching between the counting problem and the original problem. One can project the $b_{\rho,\lambda}$ onto the lattices of the grid as drawn. For example, if a path is to pass through $b_{2,4}$, it also passes through exactly one of $b_{1,3}$, $b_{2,3}$ and $b_{3,3}$. For first and third case, there is only one way to extend the path, whereas there are two ways to extend the path for the second case. (b) Schematic diagram showing how to count the total number of invalid paths. Suppose a path (blue line) is an invalid graph because it passes through the diagonal (red line). Just after it passes through the diagonal, it will reach a lattice (black dot) in another side. Reflect the remaining path (green line) by the purple line. A new path from the beginning (red dot) to the fake designation (green dot) is produced. Therefore, there is in total $C^N_{2N-2}$ invalid paths.}
\label{Fig:Proof2}
\end{figure}

This problem is equivalent to a counting problem in mathematics: suppose there is a grid of $N$ by $N$ square. Count the number of way to walk from the bottom-left corner to the top-right corner on the edges of small square with the conditions: (i) Going leftward or downward at any step is prohibited; (ii) The first step must be a step towards right; (iii) Stepping over the diagonal is prohibited. For example $N=6$, as shown in Fig. \ref{Fig:Proof2}(a). The number of steps to get to the lattice $b_{i,j}$ from the bottom-left corner satisfy the above relation (\ref{Eqn:RecursionRelationb}). The counting result is well known
\begin{equation}
b_{1,N-1} = C^{N-1}_{2N-2} - C^N_{2N-2} = \left(1- \frac{N-1}{N}\right)C^{N-1}_{2N-2} = \frac{1}{N}C^{N-1}_{2N-2}.
\end{equation}

\subsection{Higher order perturbation: random walk theory}

We proceed to study the higher order contributions by mapping the present problem to a random walk process. We shall see that this mapping has key merit in studying not only the higher order contributions, but also the phase transition induced by extra terms creating 2D and 3D fractons. To map calculation to a random walk theory, we first turn it into a probability problem. In the most general case, consider a function $f : A \rightarrow \mathbb{R}$ and suppose a summation $S$ over $A$ is wanted:
\begin{equation}
S = \sum_{a \in A} f(a)
\end{equation}
The problem can be turned into a probability question given that the size of $A$ can be calculated. Suppose $A$ is a set of events that occur with uniform probability and the reward of a particular event of $a \in A$ is $f(a)\times\left|A\right|$, the expected reward $S'$ on the set of events $A$ is
\begin{equation}
S' = \sum_{a \in A} \frac{1}{\left|A\right|} f(a)\left|A\right| = S
\end{equation}
If a non-zero normalized function $g$ (i.e. $\sum_{a\in A} g(a) = 1$) is defined on $A$, a further generalization can be made: Each particular event $a$ has $g(a)$ probability to appear with reward $f(a)/g(a)$, then
\begin{equation}
S' = \sum_{a \in A} g(a) \frac{f(a)}{g(a)} = S
\end{equation}
In the same spirit, the calculation $v_{ps}$ can be turned into a probability question:
\begin{eqnarray}
v_{ps} &=&  \sum_{p\in S_N} \frac{\left| p_{<>}\right| }{\left| S_N \right|} \frac{\left| S_N \right|}{\prod_i a_i},
\end{eqnarray}
where $\left| p_{<>}\right|/\left| S_N \right|$ can be interpreted as the probability distribution and $\left| S_N \right|/\prod_i a_i$ the reward. For higher order perturbation, a similar summation is needed, and a similar process can be made, so that the reward function is simple to find out, whereas for the probability distribution, an estimation can be made so that the size of the ultimate sum can be approximated.

Note that the calculation of the equivalent permutation splits into different stages, and each stages are independent of the other stages (c.f. the proofs recorded in last section). Therefore, the calculation of probability $\left|p_{<>}\right|/\left|S_N\right|$ can be split into products of independent probabilities, each describing how the system to jump from $a_i$ to $a_{i+1}$ with different probability to change to number of lineons. Let $L_{i,j}$ be the places that the random walker passes through, where the subscript $i$ represents the number of pairs of lineons at $j^{\rm th}$ step.
Then quantitatively,
\begin{eqnarray}
\left.P_N\right|_{L_{i,j} \rightarrow L_{i-1,j+1}} &=& \frac{i(i-1)}{(N-j)(N-j-1)} \nonumber\\
\left.P_N\right|_{L_{i,j} \rightarrow L_{i,j+1}} &=& \frac{2i(N-i-j)}{(N-j)(N-j-1)} \nonumber\\
\left.P_N\right|_{L_{i,j} \rightarrow L_{i+1,j+1}} &=& \frac{(N-i-j)(N-i-j-1)}{(N-j)(N-j-1)} \label{Eqn:RWalker},
\end{eqnarray}
where $\left.P_N\right|_{L_{i,j} \rightarrow L_{i+\delta,j+1}} $ is the probability of the random walk at $j^{\rm th}$ step go from $L_{i,j}$ to $L_{i+\delta,j+1}$ ($\delta=\pm 1,0$).

\begin{figure}[h!]
\centerline{\includegraphics[width=.8\columnwidth]{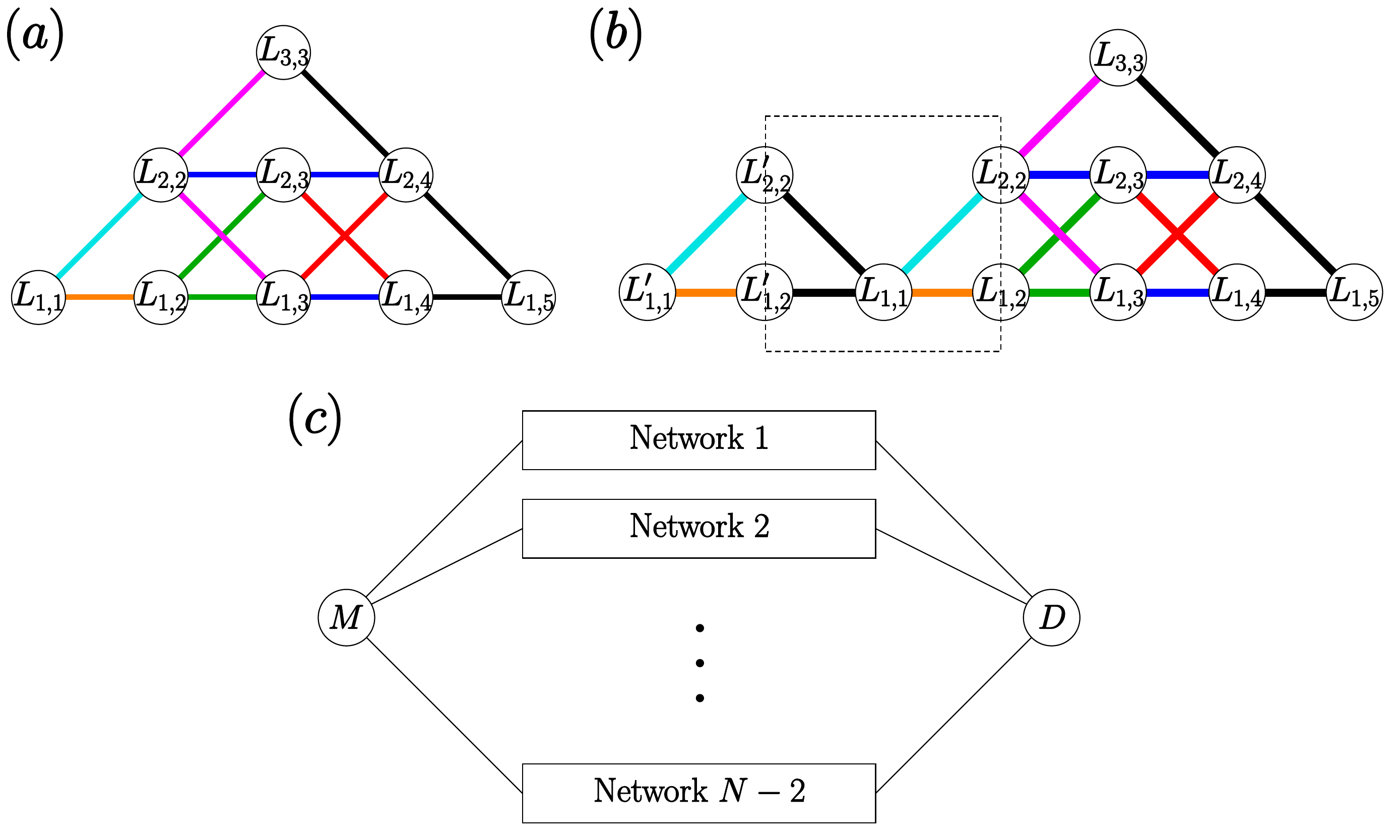}}
\caption{(a) An example of equivalent random walk for $N=6$. $L_{i,j}$ represents the $j^{\rm th}$ number in the angle notation is $i$. The walker must go rightward throughout the process. (b) An example of equivalent random walk for $r=2, N=6$. The walker is allowed to move backward once. The step that move from primed to unprimed lattice is the step when the walker move backward. The steps encircled by the black dashed rectangle are the inserted steps as described in the main passage. For both graphs, the probability for each step can be read from its color -- Cyan: $3/5$, Orange: $2/5$, Purple: $1/6$, Blue: $2/3$, Red: $1/3$ and Black: $1$. (c) Generic form of network for $r=2$. $M$ denotes the master node, from which the walker has been assigned to go to Network $i$ randomly, which is a network describing the walker must step backward after $i^{\rm th}$ step. $D$ is the dummy designation as described in the main text.}
\label{Fig:ProofH4}
\end{figure}

Consider $N=6$ (see Fig. \ref{Fig:ProofH4}(a)) for an example. Let $\alpha_i$, where $i=1,2,\cdots 6$, be the lattices on which the operator $\hat{O}$ defined. For any permutation with angle notation $\left<a_1, a_2, a_3, a_4, a_5\right>$, reinterpret it as a random walk with four steps, where probability of path taken in the $j^{\rm th}$ step represents how often the system (i) creates a pair of lineons (for $a_{j+1}-a_j = 1$), (ii) move a lineon (for $a_{j+1}-a_j = 0$), or (iii) destroy a pair (for $a_{j+1}-a_j = -1$) of lineons in the $j+1^{\rm th}$ action. Then the average reward is

\begin{eqnarray}
v_{ps} &=& \frac{1}{10} \times \frac{720}{12} + \frac{4}{15} \times \frac{720}{8} + \frac{2}{15} \times \frac{720}{4} + \frac{1}{30} \times \frac{720}{4} + \frac{1}{15} \times \frac{720}{2}\nonumber\\
&&+ \frac{2}{15} \times \frac{720}{4}+ \frac{1}{15} \times \frac{720}{2} + \frac{1}{15} \times \frac{720}{2} + \frac{2}{15} \times \frac{720}{1} = 252,
\end{eqnarray}
which is just $C^{N-1}_{2N-2}$ for $N=6$.

This current reinterpretation helps generalize the calculation to the higher order perturbation, although only estimation can be made. For the higher order perturbation, by the cluster decomposition principle, only terms with one loop are needed to be considered. They are the terms in the form $\lambda^r V_{\bar{g}g} (Q_0^{-1} V_{\bar{g}\bar{g}})^{N+r-2} Q_0^{-1} V_{g\bar{g}}$ (c.f. (\ref{Eqn:DegPerEffH})), for the $N+r^{\rm th}$ order perturbation. Let $\widetilde{S}_N^{(r)}$ be the set of the permutations of $N+r$ numbers taken from $\{1,2,\cdots, N\}$, where each number appears odd number of times. In particular, the set $S_N$ is a special case of this definition for $r=0$, i.e. $\widetilde{S}_N^{(0)} = S_N$. For each such permutation define the sum
\begin{equation}
v_{ps}^{(r)} =  \sum_{p\in \widetilde{S}_N^{(r)}} \frac{1}{\widetilde{F}_{p_1}\widetilde{F}_{p_2}\cdots\widetilde{F}_{p_{N+r-1}}} =  \sum_{p\in S_N^{(r)}} \frac{1}{\widetilde{F}_{p_1}\widetilde{F}_{p_2}\cdots\widetilde{F}_{p_{N+r-1}}}, \nonumber
\end{equation}
where $(p_1, p_2, \cdots p_{N+r-1}) \in S_N^{(r)}$, and $\widetilde{F}_{p_i}$ is interpreted as the number of pairs of lineons after applying $\prod_{j=1}^{i-1}\hat{O}_{p_j}$. Since not all permutations in $\widetilde{S}_N^{(r)}$ is valid (the $6^{\rm th}$ diagrammatic rule), a subset $S_N^{(r)}$ containing all valid permutation is considered. Then the $N+r^{\rm th}$ order term in the perturbation series can be written as $v(-v/w)^{N+r-1} v_{ps}^{(r)}$.

\subsubsection{High order terms with finite $r$}
If $r$ is a finite number, the sum $v_{ps}^{(r)}$ can be modeled by the same random walk problem, with the exception that the walker can walk backward. Take $r=2$ and $N=6$ as an example. The network $L$ is allowed to move backward once. Suppose the walker move backward at step 2 (as in Fig. \ref{Fig:ProofH4}(b)). Then it can be reinterpreted as a new network, where everywhere are the same except that an extra network (dashed boxed) is inserted. More generally, pick an element $p$ from $S_N^{(2)}$, the numbers $p_1, p_2 \cdots p_{N+2}$ are all unique except that three numbers are same. Let them be $p_\beta = p_{i+1} = p_\gamma$, where $\beta < i+1 < \gamma$. So for the first $i^{\rm th}$ actions, the effect of $i+1^{\rm th}$ and $i+2^{\rm th}$ $\hat{O}_r$ operators are to shift an applied $\hat{O}_r$ to other places. These two actions are captured by the boxed region as shown (b).

Suppose the connections and the probabilities in the boxed region has been already given exactly. The only thing left is to calculate the probability of the walker to walk backward after $i^{\rm th}$ step. Now the number of permutations $p \in S_N^{(2)}$ such that $p_{i+1}$ is the same as exactly one of $p_{1,2,\cdots,i}$ is $(i+1)(N-i) (N - 1)!$ , therefore the probability $P_i$ is
\begin{equation}
P_i = \frac{(i+1)(N-i)}{\sum_{i=1}^{N-2}(i+1)(N-i)} = \frac{6(i+1)(N-i)}{(N-2)N(N+5)}.
\end{equation}
Thus, the full network can be represented as in Fig. \ref{Fig:ProofH4}(c). The step going from the master node to the $i^{\rm th}$ network, constructed in the above method and has the probability $P_i$. At the end of the network, the walker moves to the dummy designation (which do nothing to the reward) to finish the journey. In this way, the value $v_{ps}^{(2)}$ can be exactly calculated. Although the probabilities in the boxed region can be calculated exactly, they are unimportant as $N \rightarrow \infty$. This can be seen in the following argument. (i) In this limit, nearly all walkers walk to the designation through network number $i$, such that both $i$ and $N-i$ are comparable to $N$. (Other paths have negligible probabilities to walk on.) (ii) Within these network, nearly all random walkers pass through $L_{j,i}$ such that $j$ is comparable to $N$, since the average number of segments at $i$ is of the order $N$ and the standard deviation $\sqrt{N}$. Now, since the boxed region contains 2 steps, the reward multiplier $j$ within the boxed region must lies between $j\pm 2$, which is unimportant compare to $N$. Therefore, the average reward $\overline{R}_N$ of the random walker $\overline{R}_N$ is
\begin{equation}
\overline{R}_N \approx v_{ps}\frac{\left|S_N^{(2)}\right|}{\left|S_N\right|} \int_{0}^{1} dq \frac{P(q)}{\lambda^2(q)} =  v_{ps}\left(\frac{(N+5)(N+2)}{6}\right) \int_{0}^{1} dq \frac{P(q)}{\lambda^2(q)} \approx \frac{v_{ps}}{6} \int_{0}^{1} dq \frac{P(q)}{\lambda^2(q)/N^2},
\end{equation}
where $P(q)$ is the probability density of the random walker to go backward at $qN$ step, and $\lambda(q)$ is the average number of pairs of lineons, in the original system. Since the integral is slowly varying in $N$, the ratio $\overline{R}_N / \overline{R}_{N-1}$ is still the same as $v_{ps}(N)/v_{ps}(N-1)$. Now for $r$ to be any finite even number, the arguments are still the same, except the above step are iterated finite number of times, and so the growth $\mathcal{H}_{{\rm eff},gg'}$ remains the same after including higher finite order terms. The argument can be further quantified by finding the mean and standard deviation of the number of segments. They are of order $N$ and $\sqrt{N}$, respectively, when both $i$ and $N-i$ are comparable to $N$.

Assume that the system is on open boundary condition, which is unimportant to the counting as $N \rightarrow \infty$. After $qN$ steps, suppose that there are $i$ pairs of lineons. Each pairs of lineons correspond to a line segment and the number of the configuration of the length of the segments is $S^i_{qN}$, where $S^n_r$ means the number of ways to put $r$ identical items into $n$ categories in which there are at least one item. The number of ways to distribute the remaining sites between the segments is $S^{i+1}_{N-qN+2}$. The number $2$ is present since the head and the tail can contain no unchosen segments. Since the combinatorial operator $S$ can be calculated as $S^r_n = C^{r-1}_{n-1}$,
\begin{eqnarray}
\log \lambda(q) &=& \log \frac{\sum_{i=1}^{qN} i S^i_{qN} S^{i+1}_{N-qN+2}}{C^{qN}_N} \nonumber\\
&=& \log \frac{\sum_{i=1}^{qN} i C^{i-1}_{qN-1} C^i_{N-qN+1}}{C^{qN}_N} \nonumber\\
&\approx& \frac{3}{2}\log 2\pi -\left(\frac{5}{2}+N\right)\log N+\left(2+2N(1-q)\right)\log (1-q)+2qN\log q \nonumber\\
&& + \log \int_1^{qN} \frac{di}{((i-1)!)^2(qN-i)!(N-qN-i+1)!}.
\end{eqnarray}
For the last approximate sign, Stirling's approximation is applied and terms with order smaller than $1/N$ are omitted. The integrand is sharp at its maximum and can be approximated by a Gaussian integral. The maximum can be obtained by finding the root of the derivatives of the logarithm of the integrand. Mathematically, let $x = i_{\rm max}/N$, where $i_{\rm max}$ is the place where $\lambda(q)$ is maximum:
\begin{eqnarray}
0 &=& \frac{1}{2x-2q}+N\left(\frac{1}{xN-1}-\frac{1}{2N(x+q-1)-2}\right) - N(\log N-2\log(xN-1)\nonumber\\
&&+\log(q-x)+\log(1-N(x+q-1))) \nonumber\\
x &=& \frac{-1+qN+qN^2-q^2N^2}{N(N-1)} = q-q^2,
\end{eqnarray}
where the last equal sign is correct since the third term  dominates as $N \rightarrow \infty$. Keeping the second order term, the integrand can be written as
\begin{eqnarray}
\log \int_1^{qN} \frac{di}{((i-1)!)^2(qN-i)!(N-qN-i+1)!} &\approx& \log \int_{-\infty}^\infty \exp\left(-A-B(i-x N)^2\right) di \nonumber\\
&=& -A + \frac{1}{2} \log \pi -\frac{1}{2} \log B,
\end{eqnarray}
where $A=2\log 2\pi-N+(1+N)\log N+2qN\log q+(2+2N(1-q))\log (1-q)$ and $B=N/(2(1-q)^2q^2)$. Therefore, the final result of $\lambda$ can then be acquired
\begin{equation}
\lambda(q) = q(1-q)N.
\end{equation}
The variance $\sigma(q)$ can be similarly calculated
\begin{equation}
\sigma(q) = q^2(1-q)^2N.
\end{equation}
In addition, the average number of pair of lineons $\overline{\lambda}$ and average steps $\overline{i}$ over all configuration can also be calculated.
\begin{equation}
\overline{\lambda} = \sum_{q=0;\frac{1}{N}}^1 q(1-q)N \frac{C^{\zeta N}_N}{2^N} = N^2 \int_0^1 q(1-q)\frac{C^{q N}_N}{2^N} dq = \frac{N}{4},
\end{equation}
and
\begin{equation}
\overline{i} = \sum_{q=0;\frac{1}{N}}^1 q\frac{C^{\zeta N}_N}{2^N} = \frac{N}{2},
\end{equation}
where the number after the semicolon denotes the step of the increment.

\subsubsection{High order terms with infinite $r$}
Given a random walk network for a specific $N\rightarrow \infty$. For infinite $r$, the growth of the reward as $r$ increases are dominated by the middle process that moves forward and backward a larger number of times. The result of the growth of the average reward can be proven to be $1/\overline{\lambda} = 4/N$. To capture the most essential elements, consider an one-dimensional network of random walk with sites $s_i$, $i=0,1,2,\cdots,N$ and is placed from left to right. The probability of going right is $(N-i)/N$ and left is $i/N$ for each site $i$. The probability $p_i$ of finding the random walker at $i$ after a long time satisfy
\begin{eqnarray}
p_i = p_{i-1} \frac{N-i+1}{N} + p_{i+1} \frac{i+1}{N},
\end{eqnarray}
which can be solved by $p_i = C^i_N/2^N$. Note that the probability is sharply peaked at $i_{\rm max}=N/2$. Then if the reward are written in following expansion centered at $i_{\rm max}$ - $r_i=1+\sum_j r^{(j)}(i-i_{\rm max})^j/N^j$, the expected value of the geometric mean of the rewards after walking a long time is $\prod r_i^{p_i} = 1+r^{(2)}/(4N)+O(N^{-2})$.

Indeed, the problem posted is related to a 2D random walk problem similar to Fig. \ref{Fig:ProofH4}(a), except that the random walker can move both forward or backward. The probabilities are replaced by (leading order in $N$)
\begin{eqnarray}
\left.P\right|_{L_{i,j} \rightarrow L_{i-1,j-1}} &=& \frac{i^2}{Nj} \nonumber\\
\left.P\right|_{L_{i,j} \rightarrow L_{i,j-1}} &=& \frac{2i(j-i)}{Nj} \nonumber\\
\left.P\right|_{L_{i,j} \rightarrow L_{i+1,j-1}} &=& \frac{(i-j)^2}{Nj} \nonumber\\
\left.P\right|_{L_{i,j} \rightarrow L_{i-1,j+1}} &=& \frac{i^2}{N(N-j)} \nonumber\\
\left.P\right|_{L_{i,j} \rightarrow L_{i,j+1}} &=& \frac{2i(N-i-j)}{N(N-j)} \nonumber\\
\left.P\right|_{L_{i,j} \rightarrow L_{i+1,j+1}} &=& \frac{(N-i-j)^2}{N(N-j)}.
\end{eqnarray}
The probability $p_{i,j}$ of finding the random walker at $L_{i,j}$ can be solved and is sharply peaked at $i_{\rm max}=N/4$ and $j_{\rm max}=N/2$, and the reward expanded about $N/4$ is
\begin{eqnarray}
R &=& \frac{4}{N} - \frac{16(i-N/4)}{N^2} + \frac{64(i-N/4)^2}{N^3} + O(N^{-4}) \nonumber\\
&=& \frac{4}{N}\left(1-\frac{4(i-N/4)}{N}+\frac{16(i-N/4)^2}{N^2} + O(N^{-3})\right).
\end{eqnarray}
Therefore, the geometric mean of the reward after a long time is $4/N\times(1+O(N^{-1}))$, where the coefficient of $N^{-1}$ can be found by carefully considering the probabilities and the rewards, but is unimportant in this discussion.

From here, one can see clearly that as $r$ increases by $1$, the growth of the average reward is multiplied by $4/N$. Now that the size of the set $\left|S_N^{(r=\zeta N)} \right|$ can be calculated as follows
\begin{equation}
\left|S_N^{\zeta N}\right| \approx \left|\widetilde{S}_N^{(\zeta N)} \right| \approx \frac{N^{N+\zeta N}}{2^{N-1}},
\end{equation}
which is evident when assuming the parity of the occurance of each site is independent except that the total parity matches $N(1+\zeta)$. Putting everything together, the growth of the matrix element at order $r=\zeta N$ is $N \times 4/N \times v/w = 4v/w$, so that when $v<v_c=w/4$, the matrix element is asymptotically zero. Therefore, the total perturbation can be written in the following form
\begin{equation}
\left<g\left|H_{\rm eff}\right|g'\right> = v\sum_r \mathfrak{C}_r \left(\frac{4v}{w}\right)^{N+r},
\end{equation}
where the coefficient $\mathfrak{C}_r$ increases subexponentially so that as $v< v_c = w/4$, the matrix element is zero.\\

\subsection{Energy splitting of $H_{\rm eff}$}
Let $H_{\rm eff}$ be the effective Hamiltonian of the system in the subspace of ground states, which can be written in the following form
\begin{equation}
H_{\rm eff} = \sum_r H^{(r)} \left(\frac{v}{v_c}\right)^{rN} v,
\end{equation}
where $H^{(r)}$ is a matrix independent of $v$, and the norm of its elements satisfy $\left|H_{{\rm eff},ij}^{(r)} \right|\lesssim 1$, and $N$ is the size of the system so that the total lattice sites are of the order $N^3$. Then the ground state degeneracy is of the order $2^N$ and since all ground states are on equal footing, the Hamiltonian can be written in the following form
\begin{equation}
H_{\rm eff} = \sum_{rq} h^{(r,q)} A_q \left(\frac{v}{v_c}\right)^{rN} v,
\end{equation}
where $h_G^{(r,q)} \lesssim 1$, $A_q$ are some permutation matrices, and the total number of $q$ is of the order $N^{2r}$. Therefore the maximum eigenvalue $\lambda_{H_{\rm eff}}^{\text{max}}$ of $H_{\rm eff}$ satisfies
\begin{equation}
\lambda_{H_{\rm eff}}^{\text{max}} \leq \sum_r \alpha_r N^{2r} \left(\frac{v}{v_c}\right)^{rN} v < \sum_r N^{3r}\left(\frac{v}{v_c}\right)^{rN} v,
\end{equation}
where $\alpha_r$ are constants of order $1$. This shows that $\lambda_{H_G}^{\text{max}}=0$ if $v < v_c$ and $N \rightarrow \infty$.\\

\subsection{Correction due to intersections of lineons}

The asymptotic behavior of the contribution of $\left<g\left|\mathcal{H}_{\rm eff}\right|g'\right>$, where $\left|g\right>$ and $\left|g'\right>$ are separated by multiple global line operators depends on the energy $w_i$ of two lineons from different directions meeting at a single point. Here we specifically discuss the case that $w_i < w$, where the critical point should be corrected. To model the correction, consider the case that $\left|g\right>$ and $\left|g'\right>$ are separated by a full sheet intersecting line operators. Let $\left<g\left|\mathcal{H}_{\rm eff}^{(r)}\right|g'\right> \sim \sum_\beta \prod_i F_{\beta_i}'^{-1} = \sum_\beta \prod_i \mathfrak{f}^{(r)}_{\beta,i} F_{\beta_i}^{-1}$, where $\beta$ is a configuration of successive intermediate states to connect two ground states, $F_{\beta_i}^{(')}$ are the corresponding energy in the unit $w$ after $i^{\rm th}$ action of the operators and the primed $F$ means that the energy are calculated with the assumption that lineons in different directions are independent, and $\mathfrak{f}_{\beta,i} = F_{\beta_i}/F_{\beta_i}'$. The average correction factor at order $r$ writes
\begin{equation}\tag{\ref{correctionfapprox}}
\left(\mathfrak{f}^{(r)}\right)^{N+r} = \left<\prod_i \mathfrak{f}_{\beta,i}^{(r)}\right>_\beta = \frac{\sum_\beta \prod_i^{N+r}F_{\beta_i}^{-1}\mathfrak{f}^{(r)}_{\beta,i}}{\sum_\beta \prod_i^{N+r}F_{\beta_i}^{-1}} \approx \frac{\sum_{F_{\beta_1}F_{\beta_2}\cdots} \prod_i^{N+r} M_{\beta_i} F_{\beta_i}^{-1} \mathfrak{f}^{(r)}_{F_{\beta_i}}}{\sum_{F_{\beta_1}F_{\beta_2}\cdots} \prod_i^{N+r} M_{\beta_i} F_{\beta_i}^{-1}}.
\end{equation}
As mentioned in the main text, the factor $\mathfrak{f}_{\beta,i}$ is assumed to depend on $F_{\beta_i}$ only. Note her that for sufficiently small $F_{\beta_i}$, the average number of the $F_{\beta_i}$ endpoints that meet perpendicularly is proportional to $F_{\beta_i}^2$, so the correction factor satisfies $\mathfrak{f}_{\beta_i}\sim F_{\beta_i}/\left(F_{\beta_i}-kF_{\beta_i}^2/N^2\right) \sim 1+kF_{\beta_i}/N^2$ for some constant $k$. Therefore, only when $F_{\beta_i}$ is comparable to $N^2$, the correction factor $\mathfrak{f}^{(r)}_{F_{\beta_i}}$ starts to affect the critical point. So we consider the high order terms with infinite $r$, which have largest proportion of $\beta_i$ to have $F_{\beta_i} \sim N^2$. To get a quantitative result, we first calculate $F_{\beta_i,\text{dom}}$ that dominates the contribution of $\left<g\left|\mathcal{H}_{\rm eff}^{(r=\infty)}\right|g'\right>$, and then get the corresponding average correction factors $\mathfrak{f}_{\text{dom}} = \mathfrak{f}^{(\infty)}_{F_{\beta_i,\text{dom}}}$.

We start by considering the matrix $\mathcal{M}$ that specify the random walk problem
\begin{equation}
\mathcal{M}_{ij} = \begin{cases}
\frac{x_i(x_i-\delta)}{1-\delta} & \text{for }j-i=1;\\
\frac{2x_i(1-x_i)}{1-\delta} & \text{for }j-i=0;\\
\frac{(1-x_i)(1-x_i-\delta)}{1-\delta} & \text{for }j-i=-1;\\
0 & \text{otherwise,}
\end{cases}
\end{equation}
where $x_i = 2i/(2N^2)$, $\mathcal{M}_{i+1,i(i,i/i-1,i)}$ is the probability of the system with $i$ segments to increase (remains unchange/decrease), and $\delta = 1/(2N^2)$. Therefore, the asymptotic behavior of the sum of higher order terms are captured by the matrix $\mathcal{S} = 2N^2 \mathcal{M} \odot \left[\frac{1}{\sqrt{ij}}\right]_{ij}$, where $\odot$ denotes the element-wise multiplication since a successive operations of the matrix $\mathcal{S}$ is equivalent to the calculation of the average reward of a random walk characterized by the probability matrix $\mathcal{M}$ and the reward matrix $\left[\frac{1}{\sqrt{ij}}\right]_{ij}$.  For sufficiently large $N$, the square root of the matrix $\mathcal{S}$ is
\begin{equation}
\mathcal{S}^{1/2}_{ij} = \begin{cases}
\sqrt{2}N\frac{x_i}{(ij)^{1/4}} & \text{for }j-i=-1;\\
\sqrt{2}N\frac{1-x_i}{(ij)^{1/4}} & \text{for }j-i=1;\\
0 & \text{otherwise.}
\end{cases}
\end{equation}
The eigenvector of $\mathcal{S}^{1/2}$ with largest eigenvalue describes the $F_{\beta_i}$ that dominates the contribution of the summation of high order terms with infinite $r$. Under the similarity transformation defined by a diagonal matrix $D_{ii} = \left(\prod_{j=i}^{2N^2-1} \mathcal{S}^{1/2}_{j,j+1} \left(\mathcal{S}^{1/2}_{j+1,j}\right)^{-1}\right)^{1/2}$, $\mathcal{S}^{1/2}$ becomes
\begin{equation}
\mathcal{S}'_{ij} = \left[D^{-1} \mathcal{S}^{1/2} D\right]_{ij} = \begin{cases}
\sqrt{2}N\frac{\sqrt{x_{(i+j-1)/2}\left(1-x_{(i+j+1)/2}\right)}}{(ij)^{1/4}} & \text{for }j-i=\pm 1;\\
0 & \text{otherwise.}
\end{cases}
\end{equation}
The eigenvector problem can be translated into a eigenfunction problem with a differential equation $\frac{d^2f}{dx^2} \approx 4N^4 \left(\frac{\lambda}{\sqrt{1-x}}-2\right)f$,
where $f(x)$ defines on $\left[0,1\right]$. An approximation is made by assuming $j=i$ and it is good except at $x \approx 0$, at which $j-i$ is comparable to $i$, and the true denominator of the first term in the region $x\approx 0$ should be smaller. The solution of the above differential equation can be approximated by
$f(x) \approx \text{Ai}\left(2^{2/3}N^{4/3}x+x_0\right)$, where $\text{Ai}(x)$ is the Airy function, which is the solution of the differential equation $f'' = x f$ and $x_0$ is the first zero of the Airy function. The approximation comes from the assumption that $1/\sqrt{1-x}$ is roughly linear except at large $x$. Therefore, the eigenfunction of $\mathcal{S}^{1/2}$ is
\begin{equation}
g(x) = \left(\frac{1}{\sqrt{1-x}} \left(\frac{1}{x}-1\right)^{x/2}\right)^{2N^2} f(x),
\end{equation}
where the proportional factor comes from the backward similarity transformation. As $N\rightarrow \infty$, the root of $g'(x)$ can be found by comparing coefficient of the series expansion of $g(x)$ at $N\rightarrow \infty$, which ultimately simplifies to $
\log \left(\frac{1}{x}-1\right) = 2\sqrt{x}$. The equation has a solution at $x \approx 0.26367$, so the dominating $F_{\beta_i}$ in the high order contribution with infinite $r$ is
\begin{equation}
F_{\beta_i,\text{dom}} = 0.13184 \times (2N^2).
\end{equation}
The average correction can then be calculated, by assuming that at any cross, each endpoint has $x_\text{dom} = F_\text{dom}/N^2$ probability to appear. Then the correction factor of the dominating $F_{\beta_i}$ is
\begin{equation}
\mathfrak{f}_\text{dom}^{-1} = \left(\frac{f x_{\text{dom}}^2+2x_{\text{dom}}(1-x_{\text{dom}})w/2}{wx_{\text{dom}}^2+2x_{\text{dom}}(1-x_{\text{dom}})w/2}\right) = 1-\frac{w-w_i}{w}x_\text{dom},
\end{equation}
where $w_i$ is the energy of two intersecting lineons. Therefore, the random walk calculation is enhanced by a factor $\mathfrak{f}_\text{dom}$ for each order, and so the critical point is reduced by a factor of $\mathfrak{f}_\text{dom}^{-1}$, i.e. $v=v_{c1}'=\mathfrak{f}_{\text{dom}}^{-1}\times \frac{w}{4}$. For X-cube model, $\mathfrak{f}_\text{dom}^{-1} \approx 0.87$. We estimated the errors for the X-cube model below - There are three types of errors (i) the average correction factor are calculated using arithmetic mean of $F_{\beta_i}'$ among all $\beta_i$ that have the same $F_{\beta_i}$, whereas the correct mean should be harmonic. The discrepancy can be checked for small $N$, where for $N=4$, the discrepancy is about $-0.02$ of $\mathfrak{f}_{\rm dom}^{-1}$. (ii) The approximations employed in developing and solving the different equation tend to overestimate $F_\text{dom}$. This gives a discrepancy of about $0.02$. (iii) Since the correction factor depends on $\mathfrak{f}_{\beta_i}$, the dominating $F_{\beta_i}$ is shifted when including the correction factor. This gives another $0.02$. Therefore, an estimation can be given as below $v_c = \left(0.87 \pm 0.04\right)\times w/4$.\\

\section{The upper limit of 2D and 3D fractons}
The mapping to random walk problem highly facilitate the calculation of the perturbation series as shown in section \ref{MapRW}. A similar mapping can also apply in the case of 2D or 3D fractons, where there are two important steps to identify the critical point. (i) Identify how the leading order perturbation grows as $N\rightarrow \infty$. This encodes also how the higher order perturbation with finite $r$ grows. (ii) With fixed $N$, identify how the higher order term with infinite $r$ grows. Interestingly, the sole consideration of (ii) provides a upper limit in the case of 2D and 3D fractons. It can be seen as follows: suppose the leading order term can be written in the following form
\begin{equation}
\mathcal{H}_{{\rm eff,2D},gg'}^{(N_l)} = v \beta\left(N_l\right) \left(\frac{\alpha v}{w}\right)^{N_l-1},
\end{equation}
where $N_l$ is the minimum size of a membrane operator that connects two ground states $\left|g\right>$ and $\left|g'\right>$, $\alpha$ is the growth of the leading order term as $N\rightarrow\infty$ and $\beta$ is a function of $N_l$ that is subexponential. If the growth of higher order term with infinite $r$ is known, the matrix element between the grounds state is
\begin{eqnarray}
\sum_{r} \mathcal{H}_{{\rm eff,2D},gg'}^{(N_l+r)} &=& \sum_r v \beta\left(N_l\right) \left(\frac{\alpha v}{w}\right)^{N_l-1} \beta'(N_l,r) \left(\frac{\alpha' v}{w}\right)^{r} \nonumber\\
&=& \left(\frac{\alpha}{\alpha'}\right)^{N_l}\sum_r v\beta''(N_l,r) \left(\frac{\alpha'v}{w}\right)^{N_l+r-1},
\end{eqnarray}
where $\beta'$, $\beta''$ are subexponential function of $r$, $\alpha'$ is the growth of the higher order term of infinite $r$. From here, it can be seen that as long as $r$ is sufficiently large comparing to $N$, the last term in the summation must dominate so that whenever $v>w/\alpha'=v_c^{(u)}$, the matrix element cannot be zero, and $v_c^{(u)}$ is the upper limit of the critical point. As shown in section \ref{MapRW}, the growth of the higher order term with infinite $r$ is $N/\bar{F}_d \times v/w$, where $\bar{F}_d$ is the average number of 2D ($d=2$) or 3D ($d=3$) fractons over all configuration. Therefore the upper limit for the two cases is
\begin{equation}
v_{cd}^{(u)} = \frac{\bar{F}_{d,\mathbf{r_0}}w}{2^d},
\end{equation}
where $d$ is the dimension of the fractons and $\bar{F}_{d,\mathbf{r_0}}$ is the average number of $d$d fractons at some particular point $\mathbf{r_0}$. The factor $2^d$ is present since the energy of a group of $2^d$ dD fractons is defined to be $w$.

For the case with 2D fractons, pick any point $\mathbf{r}_0$ on which a fracton can define. Let $\hat{O}_{\mathbf{r}_1}$, $\hat{O}_{\mathbf{r}_2}$, $\hat{O}_{\mathbf{r}_3}$ and $\hat{O}_{\mathbf{r}_4}$ to be the four surrouding operators listed in clockwise direction. The number of fractons at $\mathbf{r}_0$ is
\begin{equation}
F_{2,\mathbf{r}_0} = \sum_{i=1}^4\left( s_i-2s_is_{i+1}+2\prod_{j\neq i}s_j\right)-4\prod_i s_i,
\end{equation}
where $s_5=s_1$ and $s_i=(1+\hat{O}_{\mathbf{r}_i})/2$. Therefore, the average number of fractons over all possible configuration is $\bar{F}_2=3/4$ so that the upper limit of the critical point is $3w/16$.

For the case with 3D fracton, pick any point $\mathbf{r}_0$ on which a 3D fracton can define. Let $\mathbf{r}_i = \mathbf{r}_0 + l\times$ $((-1)^{i_1},(-1)^{i_2},(-1)^{i_3})$, where the binary representation of $i-1$ is $\overline{i_1i_2i_3}$ and $l$ is smallest possible length such that $\hat{O}_{\mathbf{r}_i}$ is well defined. The number of fractons at $\mathbf{r}_0$ is
 \begin{equation}
F_{3,\mathbf{r}_0} = \sum_{j=0}^{255} a_j \prod_{i=1}^8 s_i^{j_i},
\end{equation}
where the binary representation of $j$ is $\overline{j_1j_2j_3j_4j_5j_6j_7j_8}$, and the values of $a_j$'s are recorded in Table \ref{Table: 3Daj}. Therefore the average number of fractons is
\begin{equation}
\bar{F}_3 = \left(N \sum_{j=0}^{255} 2^{-\sum_i j_i} a_j\right)^{-1}=\frac{32}{29N},
\end{equation}
and the upper limit of the critical point is $256v/29w$.\\

\begin{table}[]
\begin{equation}\begin{array}{l|l}
{\rm Value} & \text{All } j \text{'s such that } a_j \text{ equals to the value} \\
\hline
\multirow{1}{*}{1} & 1,2,4,8,16,32,64,128 \\
\hline
\multirow{2}{*}{2} & 7, 11, 13, 14, 19, 21, 35, 42, 49, 50, 69, 76, 81, 84,  \\
& 112, 138, 140, 162, 168, 176, 196, 200, 208, 224 \\
\hline
\multirow{4}{*}{-2} & 3, 5, 10, 12, 17, 27, 29, 30, 34, 39, 45, 46, 48, 53, 54, 57, 58, 68, \\
& 71, 75, 78, 80, 83, 86, 89, 92, 99, 101, 106, 108, 114, 116, 120, \\
& 135, 136, 139, 141, 147, 149, 154, 156, 160, 163, 166, 169, 172, \\
& 177, 180, 184, 192, 197, 198, 201, 202, 209, 210, 216, 225, 226, 228 \\
\hline
\multirow{1}{*}{-4} & 15, 23, 43, 51, 77, 85, 113, 142, 170, 178, 204, 212, 232, 240  \\
\hline
\multirow{2}{*}{6} & 61, 62, 91, 94, 103, 110, 118, 122, 124, 155, 157, 167, \\
 & 173, 181, 185, 188, 199, 203, 211, 217, 218, 227, 229, 230 \\
\hline
\multirow{3}{*}{8} & 31, 47, 55, 59, 79, 87, 93, 107, 109, 115, 117, 121, \\
& 143, 151, 158, 171, 174, 179, 182, 186, 205, 206, \\
& 213, 214, 220, 233, 234, 236, 241, 242, 244, 248 \\
\hline
\multirow{1}{*}{-16} & 126, 189, 219, 231 \\
\hline
\multirow{2}{*}{-20} & 63, 95, 111, 119, 123, 125, 159, 175, 183, 187, 190, 207, \\
& 215, 221, 222, 235, 237, 238, 243, 245, 246, 249, 250, 252 \\
\hline
\multirow{1}{*}{44} & 127, 191, 223, 239, 247, 251, 253, 254 \\
\hline
\multirow{1}{*}{-88} & 255
\end{array} \nonumber \end{equation}
\caption{Table showing the values of $a_j$ in 3D case. The right column records all $j$ such that $a_j$ equals to the value in the corresponding cell in the left column.}
\label{Table: 3Daj}
\end{table}

\section{Lower limit of the critical points}
The main idea is to calculate the leading order perturbation $H_{\rm eff}^{(N)}$, where $N$ is the size of the system and then estimate how the matrix element grows as $N \rightarrow \infty$ to get the critical point. For lineon excitation, the leading order perturbation can be written as
\begin{equation}
\mathcal{H}_{\rm eff,gg'}^{(N)} = v\left(\frac{v}{w}\right)^{N-1} \times \sum_{p \in S_N} \frac{1}{a_p},
\end{equation}
where $S_N$ is the permutation group of order $N$, $a_p$ is the number of pairs of lineons multiplied through the process characterized by $p$. (e.g. $N=6$, $p=(1,3,5,2,4,6)$, then the numbers of pairs of lineons in each process are $1,2,3,2$ and $1$, so that $a_p=12$.). As shown above, this value can be calculated exactly:
\begin{equation}
\mathcal{H}_{\rm eff,gg'}^{(N)} = v\left(\frac{v}{w}\right)^{N-1} C^{N-1}_{2N-2} \sim v\left(\frac{v}{w}\right)^{N-1} 4^N,
\end{equation}
so that the critical point calculated in this way is
$v_c = w/4$.

\begin{figure}[h!]
\centerline{\includegraphics[width=.4\columnwidth]{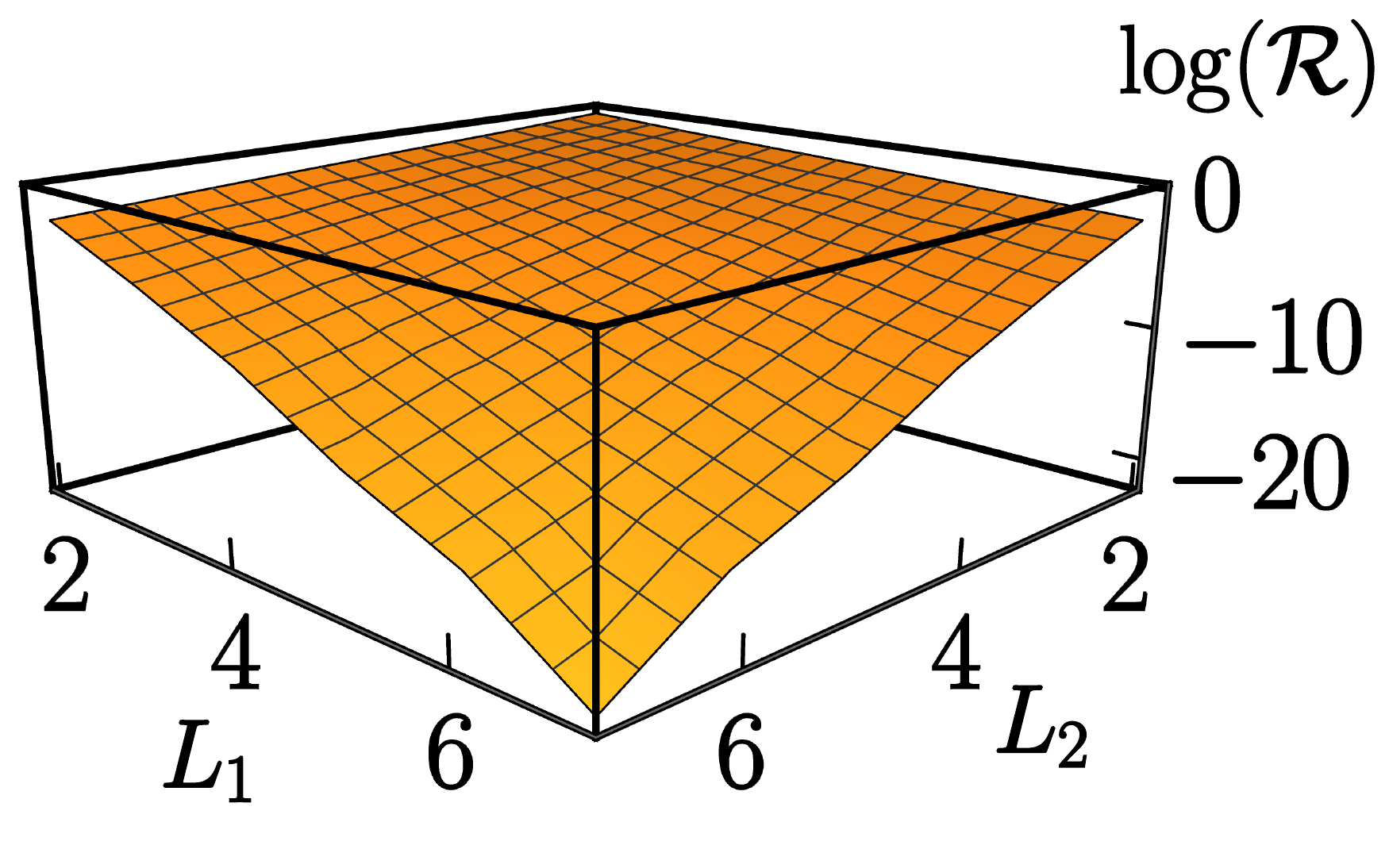}}
\caption{The graph showing the numerical evidence of $f(L_1,L_2)>F_{2,L_1L_2} (w^{L_1L_2-1}/v^N)$. $\mathcal{R}$ is the ratio of $F_{2,L_1L_2} (w^{L_1L_2-1}/v^{L_1L_2})$ to $f(L_1,L_2)$}
\label{Fig:hypnum}
\end{figure}

For fracton excitation, consider $L_1$ numbers of lines of length $L_2$ packed together. We conjecture that the growth of the matrix element is bounded by the following function
 \begin{equation}
F_{2,L_1L_2} = v\left(\frac{v}{w}\right)^{L_1L_2-1} \times \sum_{p\in S_{L_1L_2}} \frac{1}{b_p} \lesssim v\left(\frac{v}{w}\right)^{L_1L_2-1} \left( C^{L_1-1}_{2L_1-2}\right)^{L_2} \left( C^{L_2-1}_{2L_2-2}\right)^{L_1},
\end{equation}
where the notation $\lesssim$ means that the inequality is satisfied for large $N$. Write
\begin{equation}
\sum_{p\in S_{L_1L_2}} \frac{1}{b_p} = \sum_{p\in S_{L_1L_2}} \frac{1}{a_p} \frac{1}{a_p'} \frac{a_p a_p'}{b_p},
\end{equation}
where $a_p$ is the multiplication of the number of pair of lineons by treating all horizontal lines as independent, $a_p'$ is the corresponding number by treating all vertical lines as indepedent, and $b_p$ is the number of tetrad of fractons, then the last term is almost smaller than $n!$ except a few terms. In addition, we observe that $a_p$ is only weakly related to $a_p'$ so that
\begin{eqnarray}
\left(v\left(\frac{v}{w}\right)^{n-1}\right)^{-1} F_{2,L_1L_2} &\lesssim& (L_1L_2)! \sum_{p\in S_{L_1L_2}}\frac{1}{a_p} \frac{1}{a_p'} \nonumber\\
&\approx& \sum_{p \in S_{L_1L_2}} \frac{1}{a_p} \sum_{p'\in S_{L_1L_2}} \frac{1}{a_p'} \nonumber\\
&=& \left( C^{L_1-1}_{2L_1-2}\right)^{L_2} \left( C^{L_2-1}_{2L_2-2}\right)^{L_1} = f_2(L_1,L_2).
\end{eqnarray}
Numerical simulations of small $L_1$ and $L_2$ is shown in Fig. \ref{Fig:hypnum}. The ratio $\mathcal{R}$ of $F_{2,L_1L_2} (w^{L_1L_2-1}/v^{L_1L_2})$ to $f_2(L_1,L_2)$ is exponentially decaying as $L_1$ or $L_2$ increases. If the conjectured inequality is correct, when $L_1=L_2=N$, $f_2(L_1,L_2) \sim 16^{L_1L_2}=16^{N^2}$ is its asymptotic behavior. Therefore, the lower limit of the critical point for the case with 2D fractons is $v_{c2}^{l}=w/16$.

In the same spirit, we conjectured a similar inequality for 3D fractons:
\begin{equation}
F_{3,L_1L_2L_3} \lesssim v\left(\frac{v}{w}\right)^{L_1L_2L_3-1} \left(C^{L_1-1}_{2L_1-2}\right)^{L_2L_3} \left(C^{L_2-1}_{2L_2-2}\right)^{L_1L_3} \left(C^{L_3-1}_{2L_3-2}\right)^{L_1L_2} = f_3(L_1,L_2,L_3),
\end{equation}
where $L_1,L_2,L_3$ are the width, length and height of the body operator. The asymptotic behavior of $F_{3,L_1L_2L_3}$ is bounded by $f_3 \sim 64^{L_1L_2L_3}=64^{N^3}$ as $L_1=L_2=L_3=N$, so the lower limit of the critical point for the case with 3D fractons is $v_{c3}^{l}=w/64$.
\end{document}